\documentstyle[preprint,eqsecnum,prb,aps,amsfonts]{revtex}

\begin{document}

\title{
 Theory of Orbital Kondo Effect with Assisted Hopping \\
 in Strongly Correlated Electron Systems: \\
 Parquet Equations, Superconductivity and Mass Enhancement
}

\author{K. Penc}
\address{
  Institute de Physique, Universit\'e de Neuchatel\\
  1 Rue Breguet, CH-2000 Neuch\^atel, Switzerland \\
  {\rm and}\\
  Research Institute for Solid State Physics \\
  of the Hungarian Academy of Sciences  \\
  P.O.B. 49, H-1525 Budapest, Hungary
}

\author{A. Zawadowski}
\address{
  Institute of Physics,  Technical University of Budapest,\\
  XI. Budafoki \'ut 8, H-1521 Budapest, Hungary \\
  {\rm and}\\
  Research Institute for Solid State Physics \\
  of the Hungarian Academy of Sciences  \\
  P.O.B. 49, H-1525 Budapest, Hungary
}
\date{\today}

\maketitle

\begin{abstract}
 Orbital Kondo effect is treated in a model, where additional to
the conduction band there are localized orbitals close to the
Fermi energy. If the hopping between the
conduction band and the localized heavy orbitals depends on the
occupation of the atomic orbitals in the conduction band then
orbital Kondo correlation
occurs. The noncommutative nature of the coupling required for
the Kondo effect is formally due to the form factors associated
with the assisted hopping which in the momentum representation
depends on the momenta of the conduction electrons involved. The
leading logarithmic vertex corrections are due to the local
Coulomb interaction between the electrons on the heavy orbital
and in the conduction band.  The renormalized vertex functions
are obtained as a solution of a closed set of differential
equations and they show power behavior. The amplitude of large
renormalization is determined by an infrared cutoff due to finite
energy and dispersion of the heavy particles. The enhanced assisted
hopping rate results
in mass enhancement and attractive interaction in the conduction
band. The superconductivity transition temperature calculated is
largest for intermediate mass enhancement, $m^*/m \approx 2-3$.
For larger mass enhancement the
small one particle weight ($Z$) in the Green's function reduces
the transition temperature which may be characteristic for other
models as well.
 The theory is developed for different one--dimensional and square
lattice models, but the applicability is not limited to them. In
the one--dimensional case charge-- and spin--density
susceptibilities are also discussed. Good candidates for the
heavy orbital are $f$--bands in the heavy fermionic systems and
non--bonding oxygen orbitals in high temperature superconductors
and different flat bands in the quasi--one dimensional organic conductors.
\end{abstract}

\pacs{PACS numbers: 71.28.+d,72.15.Qm,74.20.-z}

\section{Introduction}
Recently the different phase transitions in strongly correlated
electronic systems has became the most intensively discussed
issue in the theory of solids.

There is a common feature in the theories of these phase
transitions, namely logarithmic corrections
characteristic for infrared divergences
occur in the perturbative expansion. That genuine character
exists even for the phase transitions in electronic systems with
the logarithmic expression $\ln(x/D)$, where $x=\max(\omega,T)$
is the largest of the energy variables $\omega$ and the
temperature $T$, furthermore, the high energy cutoff $D$ is the
bandwidth of the conduction electrons. In particular cases an
additional infrared, low energy cutoff appears which is an
inherent feature of some models. That cutoff smears out the
infrared divergences mentioned earlier.

Such logarithmic corrections occur in the theory of a
one-dimensional electron gas \cite{So79} and also e.g. in two--dimension
\cite{Dzhalo87}. In the first case that is due to the simple Fermi
surface, while in the second case either the Fermi surface is of
one--dimensional character (nested Fermi surface) or that is originated by
the corners of the Fermi surface for a nearly half--filled
electron band\cite{Dzhalo87}.
These special properties are required to get logarithmic terms in
the electron hole channel (zero sound channel), on the other
hand, it is well known from the theory of the superconductivity,
that the electron--electron channel with total zero momentum
(Cooper channel) is always divergent. The appearance of the
logarithmic contributions are generally controlled by the
conservation of momentum.
The importance of that conservation is essentially
reduced in those two--band models of arbitrary dimension, where
one of the bands is almost dispersionless thus the heavy
particle can absorb an arbitrary momentum with almost the same
energy transfer. If the heavy band is close to the Fermi energy
then the diagrams containing only one light electron (hole) in the
intermediate state contribute
by logarithmic terms arising from the integration with respect
to the energy the light electron (hole).  The finite, but small energy
and dispersion of
the heavy band introduce an inherent smearing of the logarithmic
terms, thus it represents a small infrared cutoff mentioned
earlier.

Furthermore, in order to get relevant leading order logarithmic
corrections in the absence of special features of the Fermi
surface, it is not enough to have an almost dispersionless
branch in the excitation spectrum like localized spin
excitations, two--level systems\cite{TLS} a heavy band etc., as
for the most simple models
the leading two logarithmic vertex corrections due to
intermediate states with a single light electron and
hole accompanying the heavy electron cancel each other. Such
cancellation occur in the theory
of the X--ray absorption by a deep electron level, which
phenomena was worked out by Mahan\cite{Mahan}, Nozi\`eres and de
Dominicis\cite{Xray,NodeDo69}.
The logarithmic character of single loop approximation, however, sustains
in those cases where the couplings between the light and heavy
electrons show non-trivial
structure exhibiting noncommutative behavior which may be due to
the

(i) spin dependence\cite{Kondo} (Kondo effect) or dependence on
the atomic orbital indices of a single atomic site\cite{Ba88,Cox},

(ii) dependence on the momenta of the light electrons, which
might appear as a structure factor involving two sites in the
coupling\cite{TLS,Za87,Za89PR,Za89Nob,ZaPeZi}.

The present paper is devoted to the second case, where the
structure factor is associated with the electron assisted
hopping (correlated hopping)
in a localized orbital picture\cite{Za87,Za89PR,Za89Nob,ZaPeZi}. In those
models the electron
hops between two different sites, but the hopping rate depends on
the occupation of an other site by electrons. The hopping
considered may be between two localized heavy electron states\cite{Za87} or
between a heavy and light electrons mixing the two
bands\cite{Za89PR,Za89Nob,ZaPeZi}. In
solids the mixing cases have usually larger amplitudes as the
hopping between two heavy orbitals of smaller radius are
essentially weaker.

The general form of the electron assisted interaction in the
real space contains
the following combination of creation and annihilation operators
\begin{equation}
  a^\dagger_{\alpha,n_1}
  a^{\phantom{\dagger}}_{\beta,n_2}
  a^\dagger_{\gamma,n}
  a^{\phantom{\dagger}}_{\gamma,n}
  \label{eq:genform}
\end{equation}
where $n_1$, $n_2$ are sites involved by the hopping which is
affected by the occupation of site $n$ and $\alpha$,
$\beta$, $\gamma$ are band indices. In the following only such
models are considered where either $n_1=n$ or $n_2=n$.

That interaction in the Bloch wave form is associated with a
from factors just like in the tight binding approximation.

There are two simple sources for such assisted hopping processes:

(i) the hopping between two sites must depend on the occupation
of these sites by other electrons as additional electron or
hole on the site modifies the size of the orbital in the real
space due to the Coulomb interaction, thus the hopping matrix
element is changed as well\cite{Za89PR,Za89}.

(ii) in the site representation of the two particle Coulomb
interaction which contains two creation and two annihilation
operators there are always such terms, in which one site appears
three times and another site only once\cite{Hub63,KiSuetc}.
These terms represent the off--diagonal Coulomb interaction\cite{KiSuetc}.

The form factor appearing in the Fourier transform of the operator
product given by the expression (\ref{eq:genform}) for e.g. $n_1=n$ is
\begin{equation}
  {\rm e}^{i(k_2+k_4-k_1-k_3)R_1} {\rm e}^{i k_2 \delta R}
  a^\dagger_{\alpha,k_1}
  a^{\phantom{\dagger}}_{\beta,k_2}
  a^\dagger_{\gamma,k_3}
  a^{\phantom{\dagger}}_{\gamma,k_4}
\end{equation}
where the locations of sites $n_1$ and $n_2$ are $R_1$ and
$R_2$, furthermore $\delta R=R_2-R_1$. The first factor in this
expression drops out as it ensures the momentum conservation
which is of limited relevance where one of the bands is flat.
The important form factor which is responsible for the
noncommutative nature of the interaction is the factor ${\rm
e}^{ik_2\delta R}$. Considering the light electron assisted
mixing term between the heavy and light bands one of $\alpha$
and $\beta$ belongs to the heavy band and all the others to the
light band.

The large renormalizations of the different quantities as the
strength of the interaction, the mass of the light electrons and
the electron--electron interaction relevant for
superconductivity are due to the Coulomb interaction between the
electrons in the light and heavy band which screens the charge
of the heavy electrons by the light ones. The structure of these
interaction is not crucial, thus it can be taken as a local
on--site interaction with strength $U$, except for those cases
where the model itself requires more complicated structures.
Such interactions itself without assisted hopping does not lead
to leading logarithmic corrections\cite{Ko84etc,Fukuyama}.
The theories belonging to that general class exhibit the
following common features:

(i) The vertices are renormalized and they are power functions
of the largest of the variables $\omega/D$, $T/D$ and $E_0/D$
with a low energy infrared cutoff $E_0$ discussed earlier.
In the models where there is only a structureless screening interaction
$U$ between the heavy and light electrons the leading term of
the exponent is at least quadratic in the interaction. In the case
of electron assisted band mixing treated in the present work the
exponent is linear in $U$ as that mixing interaction is
coupled to three light electrons\cite{Za89Nob}, in contrary to the
renormalization of the screening
interaction\cite{Ko84etc,Fukuyama,KaPro}. The coefficient of $U$ is different
from zero even if the structure factor is absent in the
interaction, but its presence can enhance it by a factor of two.
Such enhancement can be very important.

(ii) The light electron mass corrections exist always due to the assisted
mixing, but the large vertex corrections can lead to a large
enhancement which may reach several orders of magnitude (heavy
fermionic behavior)\cite{ZaPeZi}.

(iii) The electron--electron interaction induced by electron
assisted mixing, where the heavy electron occurs in the
intermediate state, is different from zero in the presence of
the form factor\cite{Za89PR,Za89Nob,Za89}.
The vertex corrections are important and the interaction may
depend strongly on the momenta of the electrons. Without form
factors such interactions are not generated\cite{Ko84etc,Fukuyama,KaPro}.

(iv) Both the induced electron--electron interaction and the
mass enhancement increase with the strength of coupling in the
weak coupling limit.
For stronger coupling the density of the light electron
increases, but the single particle weights Z in their spectral
functions are drastically reduced.
In the induced electron--electron interaction the square of that
weights, $Z^2$ occurs, which dominates over the increase in the
electron density which is proportional to $Z^{-1}$. Thus in
these models the strength of the superconductivity is the
strongest for moderate mass enhancement 2--3 and decreasing very
fast for larger mass enhancement ($Z\ll 1$). The largest
available dimensionless attractive coupling is close to unity.
This feature must be quite general for models where the
electron--electron interaction is induced and competes with the
mass enhancement\cite{ZaPeZi}.

In all of the behaviors listed above the energy and the
dispersion of the heavy band play determining role by limiting
the enhancement of the vertex (infrared cutoff).

It is interesting to note, that in most of the
cases\cite{Dzhalo87} of logarithmic problems new couplings are
generated by solving the
parquet or renormalization group equations for the vertex
functions. That is the case for the two dimensional electron gas
as well\cite{Dzhalo87}. In the present case the finite number of
the generated form factors reduces the problem to a closed set
of differential equations, and their solution will be presented
in analytical form.

The models where the present theory can be relevant are those,
where there is a heavy flat band. Those bands can be formed by
the atomic orbitals of very small size like $4f$ electrons in
heavy fermionic systems and the mixing is between $s-$ and
$f-$electrons. Another possibility is the case of non--bonding
orbitals in complicated crystal structures\cite{Za89}, where these orbitals
are only weakly hybridized with the conducting electrons, e.g.
non--bonding $\pi-$orbital of oxygen in the high temperature
superconductors or some appropriate orbitals in the quasi
one--dimensional conducting molecular crystals where there might
be several flat bands.
The present paper is not devoted to study particular special
cases, but to provide several methods which can be applied to
different concrete cases.

The paper is organized as follows: In chapter II. we are
presenting the Hamiltonian of a general model with features
described above. In chapter III. we are calculating the vertex
corrections for the general model. In chapters IV.
and V. the calculations of the self--energy and the
electron--electron interaction are presented and in the next
chapter (VI.) the superconducting transition temperature is
determined. That discussion of the relation of the mass
enhancement to the generated electron--electron interaction
contains quite general considerations which are valid much
beyond the model treated here. The formalism developed in the
previous chapters is
applied to special cases: to the one dimensional electron gas
(chapter VII., see also Ref. \onlinecite{Za89PR}) and to
two models which has a resemblance to the $CuO_2$ plane of
the high temperature superconductors
(chapter VIII.). In the latter case two simplified models are
presented where the solution of vertex equations are essentially
different. Finally, in chapter IX. we give a brief comprehensive
conclusion with some hint concerning the applicability of the
model. The chapters VII. and VIII. can be read independently.

\section{THE MODEL}

 The model proposed consists of two electron bands in the tight
binding approximation: the broad light ($l$) band with bandwidth
$D$ and the narrow heavy ($h$) one with energy $\varepsilon_h$
($|\varepsilon_h|\ll D$) measured from the Fermi energy
$\varepsilon_F^{\vphantom{\dagger}}$.
 The $l$-orbitals are on each atom.
 The heavy orbitals are located at some of the atoms in the cell and
they overlap only with $l$-orbitals of the nearest atoms.
In general theory they may be more than one heavy orbital at one site,
 which corresponds to index
$\gamma$. In Fig.~\ref{fig:1Dmod} we are showing such a model in one
dimensions, where the light orbitals are $s$ orbitals and
and the heavy orbitals are of $p$ and $d$ type. In Fig.~\ref{fig:2Dmod}
as an example of a two dimensional model, the $CuO_2$ plane and the
apical oxygens below the $Cu$ are shown.
The heavy orbitals are associated with
the $p_x$ and $p_y$ orbitals of the apical oxygen.

The hopping Hamiltonian   can be given in the terms of the
annihilation operators $c_{{\bf n}{\bbox{\delta}}\sigma}$, and
$h_{{\bf n}\gamma\sigma}$, where ${\bf n}$ stands for the
position of the atom with heavy orbitals, the spin is
$\sigma=\pm1$ . The (${\bf n},{\bbox{\delta}}$) labels those light
atoms, which are the first neighbors of the site ${\bf n}$ at
$({\bf n} + {\bbox{\delta}}/2)a$, or at site ${\bf n}$ for ${\bbox{\delta}}=0$,
 where $a$ is the lattice constant.
The labels of the $c$-operators are not defined in
a unique way. For example in the case of square lattice both
(${\bf n},{\bbox{\delta}}$) and (${\bf n}+{\bbox{\delta}},-{\bbox{\delta}}$)
labels the same atom on the positions
$({\bf n} + {\bbox{\delta}}/2)a$. Thus the one-particle Hamiltonian $H_0$ is
\begin{eqnarray}
H_0 & = &  \sum_{{\bf n}, \sigma}
   (\varepsilon_{\bbox{\delta}}+\varepsilon^{\vphantom{\dagger}}_F)
    n^{\phantom{\dagger}}_{{\bf n}{\bbox{\delta}}\sigma}
  + (\varepsilon_h^{\vphantom{\dagger}}+\varepsilon_F^{\vphantom{\dagger}})
    \sum_{{\bf n}, \sigma, \gamma}
    h^\dagger_{{\bf n}\gamma\sigma}
    h^{\phantom{\dagger}}_{{\bf n}\gamma\sigma}
  \nonumber \\
  & + &
    \sum_{{\bf n}, \sigma,{\bbox{\delta}},\gamma}
    t^\gamma_{h{\bbox{\delta}}}
    h^\dagger_{{\bf n}\gamma\sigma}
    c^{\vphantom{\dagger}}_{{\bf n}{\bbox{\delta}}\sigma}
  + \sum_{{\bf n}, \sigma, {\bbox{\delta}},{\bbox{\delta}}'}
    t^{\vphantom{\dagger}}_{{\bbox{\delta}}{\bbox{\delta}}'}
    c^\dagger_{{\bf n}{\bbox{\delta}}\sigma}
    c^{\vphantom{\dagger}}_{{\bf n}{\bbox{\delta}}'\sigma}
  +{\rm h.c.} \;,  \label{eq:H0}
\end{eqnarray}
where $\varepsilon_{\bbox{\delta}}$ is an energy splitting,
$t_{{\bbox{\delta}}{\bbox{\delta}}'}$ and $t^\gamma_{h,\bbox{\delta}}$
 are hopping parameters
($t^\gamma_{h,\bbox{\delta}=0} \ll
t^{\vphantom{\dagger}}_{{\bbox{\delta}}{\bbox{\delta}}'}$) and
the definition $n^{\vphantom{\dagger}}_{{\bf n}{\bbox{\delta}}\sigma}=
c^\dagger_{{\bf n}{\bbox{\delta}}\sigma} c^{\vphantom{\dagger}}_{{\bf
n}{\bbox{\delta}}\sigma} $
 is used.
$t^\gamma_{h,\bbox{\delta}}$ must hold because of the different symmetry of
the orbitals. The hopping $t_{{\bbox{\delta}}{\bbox{\delta}}'}$ in the light
band may include the site ${\bbox{\delta}}=0$ as well.
The direct weak hoppings between the light and heavy bands will be taken into
 account as a broadening of the $h$-orbitals ($\Gamma$), which may serve as a
low energy cutoff in the logarithmic integrals.
The part of Eq.~(\ref{eq:H0}) due to the $l$-orbitals can be
diagonalized and only the band crossing the Fermi  energy is
kept. In the new band the annihilation operator is denoted by
$d_{{\bf k}\sigma}$ where ${\bf k}$ is the momentum. The
contributions to the
Fourier transforms of the $c_{{\bf k}{\bbox{\delta}}\sigma}$ of the $d$-band
crossing the Fermi surface are
\begin{equation}
   c_{{\bf k }\bbox{\delta}\sigma}
   =\phi_{\bbox{\delta}}({\bf k})d_{{\bf k}\sigma} \;,
  \label{eq:a=phi_d}
\end{equation}
where $\phi_{{\bbox{\delta}}}({\bf k})$ are the amplitudes of
the electron on the orbital denoted by ${\bbox{\delta}}$ in the state
${\bf k}$ and the contributions of the other bands are dropped.

The interaction Hamiltonian consists of two parts,
$H_{\rm int}=H_U+ H_{\tilde t}$. The Hamiltonian describing the Coulomb
repulsion is given by
\begin{equation}
  H_U =  \sum_{{\bf n},{\bbox{\delta}}}
    \sum_{\gamma,\sigma,\sigma'}
    U^\gamma_{{\bbox{\delta}}}
    h^\dagger_{{\bf n}\gamma\sigma\vphantom{{\bbox{\delta}}}}
    c^\dagger_{{\bf n}{\bbox{\delta}}\sigma'}
    c^{\vphantom{\dagger}}_{{\bf n}{\bbox{\delta}}\sigma'}
    h^{\vphantom{\dagger}}_{{\bf n}\gamma\sigma\vphantom{{\bbox{\delta}}}}
    + {\rm h.c.} \;,
  \label{eq:HU}
\end{equation}
where  $U^\gamma_{{\bbox{\delta}}}$ is the Coulomb integral between
the heavy orbital $\gamma$ and light orbitals at site
${\bbox{\delta}}$ (see e.g. Fig.~\ref{fig:1Dmod}c).
 The Hamiltonian due to assisted hopping is
\begin{eqnarray}
 H_{\tilde t} &=&
  \sum_{{\bf n},{\bbox{\delta}}} \sum_{\gamma,\sigma}
  \tilde t^\gamma_{{\bbox{\delta}}}
  n^{\vphantom{\dagger}}_{{\bf n}{\bbox{\delta}}-\sigma}
  c^\dagger_{{\bf n}{\bbox{\delta}}\sigma}
  h^{\vphantom{\dagger}}_{{\bf n}\gamma\sigma\vphantom{{\bbox{\delta}}}}
 + {\rm h.c.} \nonumber\\
 &=&
 \sum_{{\bf n},{\bbox{\delta}}} \sum_{\gamma,\sigma}
 \tilde t^\gamma_{{\bbox{\delta}}}
  c^\dagger_{{\bf n}{\bbox{\delta}}\sigma}
  c^\dagger_{{\bf n}{\bbox{\delta}}-\sigma}
  c^{\vphantom{\dagger}}_{{\bf n}{\bbox{\delta}}-\sigma}
  h^{\vphantom{\dagger}}_{{\bf n}\gamma\sigma\vphantom{{\bbox{\delta}}}}
  + {\rm
h.c.} \;,
\label{eq:Hasst}
\end{eqnarray}
where $\tilde t_{\bbox{\delta}}^\gamma$ is the amplitude of the assisted
hopping between the heavy
orbital $\gamma$ and light orbitals at site ${\bbox{\delta}}$.  The
possible role of the on-site correlations between the heavy
electrons are taken into account in the renormalized parameter
$\varepsilon_h$ which can be used if only the single excitation of
smallest energy is considered at each $h$-site and the other
excitations shifted in energy due to the large on--site Coulomb
interaction are dropped.

The form factor responsible for the orbital Kondo-effect arises
from the Fourier transform of the Hamiltonian $H_{\tilde t}$
describing the assisted hopping,
\begin{eqnarray}
  H_{\tilde t} = { 1 \over N^2} \sum_{{\bf k}_1,{\bf k}_2,{\bf k}_3}
  \sum_{{\bbox{\delta}},\gamma,\sigma }
  && \tilde t^\gamma_{{\bbox{\delta}}}
  {\rm e}^{-i({\bf k}_1+{\bf k}_2 -{\bf k}_3){\bbox{\delta}} a/2}
  \phi^*_{{\bbox{\delta}}}({\bf k}_1)
  \phi^*_{{\bbox{\delta}}}({\bf k}_2)
  \phi_{{\bbox{\delta}}}({\bf k}_3) \nonumber\\
  &&\times
  d^\dagger_{{\bf k}_1\sigma}
  d^\dagger_{{\bf k}_2-\sigma}
  d^{\vphantom{\dagger}}_{{\bf k}_3-\sigma}
  h^{\vphantom{\dagger}}_{\gamma{\bf k}_h\sigma} + {\rm {}h.c.} \;,
  \label{eq:Hasst_k}
\end{eqnarray}
where ${\bf k}_h={\bf k}_1+{\bf k}_2-{\bf k}_3$, the notation
given by Eq.~(\ref{eq:a=phi_d}) is introduced  and $N$ is the number  of
 unit cells in the sample. This form factors prevents the cancellation of the
logarithmically divergent loops which usually cancel if the interaction terms
are diagonal. In Fig.~\ref{fig:bare_tU} the diagrammatic representation of the
bare interaction is shown.

Along the calculation several approximations will be applied.
The momentum integrals are split to an energy integral
perpendicular to the Fermi surface and a momentum integral on
the Fermi surface. Thus e.g. in $d$-dimensions
\begin{equation}
  {1 \over N} \sum_k = a^d \int{ d^d {\bf k} \over (2\pi)^d}
  \quad\rightarrow\quad
  {a^d\over (2\pi)^d}
  \int_S {dS_{\bbox{\kappa}} \over v^{\vphantom{\dagger}}_F(\bbox{\kappa})}
  \int^D_{-D}d\varepsilon \;,
  \label{eq:sum2int}
\end{equation}
where $dS_{\bbox{\kappa}}$ is a surface element of the Fermi
surface, $\bbox{\kappa}$ is the unit vector directed from the center
of the Brillouin zone to the surface element $dS$, and ${\bf
k}^{\vphantom{\dagger}}_F(\bbox{\kappa})$ is the Fermi wave vector in the
direction $\bbox{\kappa}$. Furthermore, the form factor appearing in
Eq.~(\ref{eq:Hasst_k}) and the wave functions
$\phi_{{\bbox{\delta}}}({\bf k})$ are replaced by their values taken
at the Fermi surface like
\begin{equation}
  {\rm e}^{\displaystyle -i({\bf k}_1+{\bf k}_2 -{\bf k}_3){\bbox{\delta}} a/2}
  \quad\rightarrow\quad
  {\rm e}^{\displaystyle -i[ {\bf k}^{\vphantom{\dagger}}_F(\bbox{\kappa}_1)
                           +{\bf k}^{\vphantom{\dagger}}_F(\bbox{\kappa}_2)
                           -{\bf k}^{\vphantom{\dagger}}_F(\bbox{\kappa}_3)]
           {\bbox{\delta}} a/2} \;'
  \label{eq:expFS}
\end{equation}
and
\begin{equation}
  \phi_{{\bbox{\delta}}}({\bf k})\quad\rightarrow\quad
  \phi_{{\bbox{\delta}}}({\bf k})\vert_{{\bf k} =
  {\bf k}^{\vphantom{\dagger}}_F(\bbox{\kappa})}
    =\phi_{{\bbox{\delta}}}(\bbox{\kappa})\;.
  \label{eq:phiFS}
\end{equation}
The Fermi velocity $v^{\vphantom{\dagger}}_F(\bbox{\kappa})$ is defined as
\begin{equation}
  \left \vert
    \partial\varepsilon\over \partial {\bf k}
  \right \vert_{{\bf k}={\bf k}^{\vphantom{\dagger}}_F(\bbox{\kappa})}
  =v^{\vphantom{\dagger}}_F(\bbox{\kappa})\;.
  \label{eq:vF}
\end{equation}

\section{VERTEX CORRECTIONS}

The renormalization scheme consists of two steps due to:

(i) Coulomb interaction $H_U$ which in the  leading logarithmic
approximation results in the corrections $\tilde t U^n
\ln^n|D/\varepsilon|$, (n=1,2,3...) to the vertex $\tilde t$ (see
Fig.~\ref{fig:corr}), where the energy cutoff $\varepsilon$ is
the largest of the energy variables, $\varepsilon={\rm
Max}(|\omega|,|\varepsilon_h|,\Gamma)$;

(ii) assisted hopping $H_{\tilde t}$ in the
 next to leading logarithmic approximation by calculating the
self--energies proportional to $\omega\tilde t^2 \ln D/\varepsilon$ and
the $l$-particle four vertex function $\sim \tilde t^2 \ln
D/\varepsilon$. As a consequence an identity of Ward type, the self--energy
depicted on Fig.~\ref{fig:self_ene} is connected with the vertex
correction giving the first contribution to  the induced
interaction $V$ in the electron-hole
channel (see Sec. V).

 In both steps the $h$-orbitals at different sites contribute
independently.  Joint contributions of different sites are relevant
only in higher logarithmic orders which are neglected.
 The intermediate states with two $l$-particles or two holes do
not
contribute to the  renormalization as logarithmic contribution
arises from the particle-particle channel only if their
total momentum is zero.  The latter is the case of the
superconducting gap equation, but not in the processes described
above.

The strong renormalization of $\tilde t$ is generated by three
diagrams which are depicted in Fig.~\ref{fig:corr}b. The
summation of these diagrams are performed in the leading
logarithmic approximation considering the ``parquet" diagrams. As
$\tilde t$ is a small variable compared with $U$ $(|\tilde
t|\ll|U|)$, the equation are linearized in $\tilde t$; thus two or more
$h$-particle intermediate states are not included. This
approximation is consistent as the parquet scheme does not
contain these diagrams. The parquet equation for vertex $\tilde
t$ can be written in a schematic form as
\begin{equation}
  \tilde t(\omega)=\tilde t + \cdots
  \int^D_{{\rm Max}(|\omega|,|\varepsilon_h|)}
  \tilde t(\varepsilon){1\over (-\varepsilon)} U(\varepsilon) d\varepsilon\;,
  \label{eq:sca_int}
\end{equation}
where the momentum integrals and the coefficient are not presented
and $U (\omega)=U$ must be taken as it will be discussed later.
 The variable $\varepsilon$ represents the
smallest energy denominator in the one $h$-particle and one
$d$-particle (hole) channel.

The differential form of the schematic Eq.~(\ref{eq:sca_int}) is the scaling
equation
\begin{equation}
  \omega{\partial \tilde t(\omega) \over \partial\omega}
  =\cdots\tilde t (\omega)U\;.
  \label{eq:sca_diff}
\end{equation}

The renormalization procedure generates an effective
assisted hopping, which can be given as
\begin{equation}
  {1 \over N^2}\sum_{\gamma,\sigma}
  \sum_{{\bf k}_1,{\bf k}_2,{\bf k}_3}
    \tilde
t^\gamma(\bbox{\kappa}_1,\bbox{\kappa}_2,\bbox{\kappa}_3;\varepsilon)
    d^\dagger_{{\bf k}_1\sigma}
    d^\dagger_{{\bf k}_2-\sigma}
    d^{\vphantom{\dagger}}_{{\bf k}_3-\sigma}
    h^{\vphantom{\dagger}}_{{\bf k}_h\gamma\sigma}\;,
  \label{eq:effasst}
\end{equation}
where ${\bf k}_h={\bf k}_1+{\bf k}_2-{\bf k}_3$. The
unrenormalized hopping [see Eq.~(\ref{eq:Hasst_k})] is
\begin{equation}
  \tilde t^{\gamma(0)}(\bbox{\kappa}_1,\bbox{\kappa}_2,\bbox{\kappa}_3)=
    \sum_{{\bbox{\delta}}}
    \tilde t^\gamma_{{\bbox{\delta}}}
    {\rm e}^{-i\bigr[
      {\bf k}_F(\bbox{\kappa}_1)+{\bf k}_F(\bbox{\kappa}_2)
      -{\bf k}_F(\bbox{\kappa}_3)\bigl]{\bbox{\delta}} a/2}
    \phi^*_{{\bbox{\delta}}}(\bbox{\kappa}_1)
    \phi^*_{{\bbox{\delta}}}(\bbox{\kappa}_2)
    \phi_{{\bbox{\delta}}}(\bbox{\kappa}_3)
  \label{eq:asst_unren} \;.
\end{equation}
Similarly, the momentum representation of the effective Coulomb repulsion
is
\begin{equation}
  {1 \over N^2}\sum_{\gamma,\sigma}
  \sum_{{\bf k}_1,{\bf k}_2,{\bf k}_{h1}}
    U^\gamma(\bbox{\kappa}_1,\bbox{\kappa}_2)
    h^\dagger_{{\bf k}_{h1}\gamma\sigma}
    d^\dagger_{{\bf k}_1-\sigma}
    d^{\vphantom{\dagger}}_{{\bf k}_2-\sigma}
    h^{\vphantom{\dagger}}_{{\bf k}_{h2}\gamma\sigma} \;.
  \label{eq:HU_k}
\end{equation}
where ${\bf k}_{h2}={\bf k}_{h1}+{\bf k}_1-{\bf k}_2$.

The present treatment is essentially simplified by the fact,
that logarithmic vertex correction of the Coulomb vertices shown
in Fig.~\ref{fig:corr}a cancel out as they describe static
potential represented by the $h$-particle, thus the scattering
strength $U^\gamma_{{\bbox{\delta}}}$ is not renormalized.
   Thus the effective Coulomb interaction
$U^\gamma(\bbox{\kappa},\bbox{\kappa}')$ remains diagonal and can be given
as
\begin{equation}
    U^\gamma(\bbox{\kappa},\bbox{\kappa}') =
    \sum_{\bbox{\delta}} U^\gamma_{{\bbox{\delta}}}
     {\rm e}^{
       -i[{\bf k}^{\vphantom{\dagger}}_F(\bbox{\kappa})
         -{\bf k}^{\vphantom{\dagger}}_F(\bbox{\kappa}')]
       {\bbox{\delta}} a/2}
    \phi^*_{\bbox{\delta}}(\bbox{\kappa})
    \phi_{\bbox{\delta}}(\bbox{\kappa}')
 \label{eq:Ukk}\;.
\end{equation}

In the case of the assisted hopping such cancelations
does not occur, since the form
factors in front of the loop contributions are not the same,
and we can get the following scaling
equation in leading logarithmic order:
\begin{eqnarray}
  \omega
  { \partial  \over \partial \omega} \tilde
     t^\gamma(\bbox{\kappa}_1,\bbox{\kappa}_2,\bbox{\kappa}_3,\omega)
  & = & \rho\bigl\langle
      U^\gamma(\bbox{\kappa}_1,\bbox{\kappa})
      \tilde
   t^\gamma(\bbox{\kappa},\bbox{\kappa}_2,\bbox{\kappa}_3,\omega)
    \bigr\rangle_{\bbox{\kappa}} \nonumber\\
 & + & \rho\bigl\langle
      U^\gamma(\bbox{\kappa}_2,\bbox{\kappa})
      \tilde
    t^\gamma(\bbox{\kappa}_1,\bbox{\kappa},\bbox{\kappa}_3,\omega)
    \bigr\rangle_{\bbox{\kappa}} \nonumber\\
  & - & \rho\bigl\langle
      \tilde
    t^\gamma(\bbox{\kappa}_1,\bbox{\kappa}_2,\bbox{\kappa},\omega)
      U^\gamma(\bbox{\kappa},\bbox{\kappa}_3)
    \bigr\rangle_{\bbox{\kappa}} \label{eq:sca_tk1k2k3} \;,
\end{eqnarray}
where the three terms on the right hand side corresponds to diagrams
 shown in Fig.~\ref{fig:corr}b.
Here we introduced the notation of the average of a function
$f(\bbox{\kappa})$
 over the Fermi surface as
 \begin{equation}
   \bigl\langle
     f(\bbox{\kappa})
   \bigr\rangle_{\bbox{\kappa}} = {a^d \over (2\pi)^d \rho}
  \int { dS_{\bbox{\kappa}} \over v^{\vphantom{\dagger}}_F(\bbox{\kappa})}
f(\bbox{\kappa})
  \;,
  \label{eq:def_ave}
\end{equation}
in dimension $d$, and $\rho$ is the density of states,
\begin{equation}
  \rho= {a^d \over (2\pi)^d }
      \int { dS_{\bbox{\kappa}} \over
v^{\vphantom{\dagger}}_F(\bbox{\kappa})} \;.
  \label{eq:rho}
\end{equation}

 The scaling Eq~.(\ref{eq:sca_tk1k2k3}) can be transformed to a  real
space representation by assuming the Ansatz
\begin{eqnarray}
 \tilde
  t^\gamma(\bbox{\kappa}_1,\bbox{\kappa}_2,\bbox{\kappa}_3,\omega)
  = \sum_{{\bbox{\delta}}_1,{\bbox{\delta}}_2,{\bbox{\delta}}_3}
   &&\tilde t^\gamma_{{\bbox{\delta}}_1{\bbox{\delta}}_2{\bbox{\delta}}_3}
  (\omega)
  \phi^*_{{\bbox{\delta}}_1}(\bbox{\kappa}_1)
  \phi^*_{{\bbox{\delta}}_2}(\bbox{\kappa}_2)
  \phi_{{\bbox{\delta}}_3}(\bbox{\kappa}_3) \nonumber\\
  &&\times{\rm e}^{
      -i[ {\bf k}^{\vphantom{\dagger}}_F(\bbox{\kappa}_1){\bbox{\delta}}_1
        + {\bf k}^{\vphantom{\dagger}}_F(\bbox{\kappa}_2){\bbox{\delta}}_2
        - {\bf k}^{\vphantom{\dagger}}_F(\bbox{\kappa}_3){\bbox{\delta}}_3
        ] a/2}
  \label{eq:def_tddd} \;,
\end{eqnarray}
(where ${\bbox{\delta}}$ points to nearest neighbor or ${\bbox{\delta}}=0$),
thus the following closed set of equations is obtained:
\begin{equation}
  \omega
  { \partial \tilde
  t^\gamma_{{\bbox{\delta}}_1{\bbox{\delta}}_2{\bbox{\delta}}_3}(\omega)
     \over \partial \omega}
  =  \rho \sum_{{\bbox{\delta}}}
     \Bigl(
      U^\gamma_{{\bbox{\delta}}_1} F_{{\bbox{\delta}}_1{\bbox{\delta}}}
      \tilde
   t^\gamma_{{\bbox{\delta}}{\bbox{\delta}}_2{\bbox{\delta}}_3}(\omega)
   +  U^\gamma_{{\bbox{\delta}}_2} F_{{\bbox{\delta}}_2{\bbox{\delta}}}
      \tilde
   t^\gamma_{{\bbox{\delta}}_1{\bbox{\delta}}{\bbox{\delta}}_3}(\omega)
   -  \tilde
   t^\gamma_{{\bbox{\delta}}_1{\bbox{\delta}}_2{\bbox{\delta}}}(\omega)
      F_{{\bbox{\delta}}{\bbox{\delta}}_3} U^\gamma_{{\bbox{\delta}}_3}
     \Bigr)\;.
  \label{eq:sca_tddd}
\end{equation}
Here the $F_{{\bbox{\delta}}{\bbox{\delta}}'}$ incorporates the form
factors by the definition
\begin{equation}
  F_{{\bbox{\delta}}{\bbox{\delta}}'}
  = \bigl\langle
     {\rm e}^{
       i{\bf k}^{\vphantom{\dagger}}_F(\bbox{\kappa})
      (\bbox{\delta}-\bbox{\delta}')a/2}
    \phi_{\bbox{\delta}}(\bbox{\kappa})
    \phi^*_{{\bbox{\delta}}'}(\bbox{\kappa})
  \bigr\rangle
 \label{eq:Fdd}
\end{equation}
which appears in the transition amplitude of an $l$--electron between
two sites, ${\bbox{\delta}}$ and ${\bbox{\delta}}'$, in the tight binding
approximation.
 As a special case, the $\rho_{{\bbox{\delta}}}=\rho
F_{{\bbox{\delta}}{\bbox{\delta}}}$ gives the partial density of states at a
site ${\bbox{\delta}}$.

The solution of the differential Eq.~(\ref{eq:sca_tddd}) can be
given immediately as:
\FL
\begin{equation}
  \tilde t^\gamma_{{\bbox{\delta}}_1{\bbox{\delta}}_2{\bbox{\delta}}_3}(\omega)
  =  \sum_{{\bbox{\delta}}_1'{\bbox{\delta}}_2'{\bbox{\delta}}_3'}
     \left[
       \left( D \over \omega \right)^{\rho {\sf U}^\gamma{\sf F}}
     \right]_{{\bbox{\delta}}_1{\bbox{\delta}}_1'}
     \left[
       \left( D \over \omega \right)^{\rho {\sf U}^\gamma{\sf F}}
     \right]_{{\bbox{\delta}}_2{\bbox{\delta}}_2'}
     \tilde
t^{\gamma(0)}_{{\bbox{\delta}}_1'{\bbox{\delta}}_2'{\bbox{\delta}}_3'}
     \left[
       \left( D\over \omega \right)^{-\rho {\sf F}{\sf U}^\gamma}
     \right]_{{\bbox{\delta}}_3'{\bbox{\delta}}_3} \;,
\label{eq:sol_tddd}
\end{equation}
where we used the notation ${\sf F}$ and ${\sf U}^\gamma$ for the matrix
$F_{{\bbox{\delta}}{\bbox{\delta}}'}$ and the diagonal matrix
$U^\gamma_{{\bbox{\delta}}{\bbox{\delta}}'}=U^\gamma_{{\bbox{\delta}}}
\delta_{{\bbox{\delta}}{\bbox{\delta}}'}$.
Unfortunately this form of the solution is not suitable for
further calculations. In the actual computation let us consider the
eigenvalue problem of the matrix ${\sf U}^\gamma{\sf F}$
appearing in the exponent:
\begin{equation}
  \sum_{{\bbox{\delta}}'}
  U^\gamma_{{\bbox{\delta}}} F_{{\bbox{\delta}}{\bbox{\delta}}'}
  s^{\gamma,j}_{{\bbox{\delta}}'} = \lambda^\gamma_j
  s^{\gamma,j}_{{\bbox{\delta}}} \label{eq:eigenFU} \;,
\end{equation}
where $j$ labels the eigenvalues thus the spectral decomposition of the
matrix ${\sf U^\gamma F}$, which
is not necessarily uniter, can be given as
\begin{equation}
   U^\gamma_{{\bbox{\delta}}} F_{{\bbox{\delta}}{\bbox{\delta}}'}  =
   \sum_j \lambda_j^\gamma  s^{\gamma,j}_{{\bbox{\delta}}}
   r^{\gamma,j}_{{\bbox{\delta}}'} \;,
  \label{eq:spectFU}
\end{equation}
with the row vectors ${\bf r}$ orthogonal to the column
vectors ${\bf s}$, thus
\begin{equation}
  \sum_{\bbox{\delta}} r_{\bbox{\delta}}^{\gamma,j}
    s_{\bbox{\delta}}^{\gamma,j'} = \delta_{jj'} \;.
  \label{eq:ortho_rs}
\end{equation}
In general the matrix ${\sf U^\gamma F}$ is not invertable, that is why the
left and the right eigenvectors can be different, such an example is presented
in Sec. VIII. (case A).

Using the identity
$({\sf U}^\gamma {\sf F})_{{\bbox{\delta}}{\bbox{\delta}}'}=
 ( {\sf F U}^\gamma)^*_{{\bbox{\delta}}'{\bbox{\delta}}}$, which
 follows form Eq.~(\ref{eq:Fdd}) and that ${\sf U}^\gamma$ is
diagonal and real, thus the left eigenvectors of
$ {\sf F U}^\gamma$ are the $s^{\gamma,j*}$
and writing
$\tilde t^\gamma_{{\bbox{\delta}}_1{\bbox{\delta}}_2{\bbox{\delta}}_3}$ as a
linear combination of the eigenvectors ${\bf s}$ as
\begin{equation}
   \tilde t^\gamma_{{\bbox{\delta}}_1{\bbox{\delta}}_2{\bbox{\delta}}_3} =
   \sum_{ijk} \tilde t^\gamma_{ijk}(\omega)
    s^{\gamma,i}_{{\bbox{\delta}}_1} s^{\gamma,j}_{{\bbox{\delta}}_2}
    s^{*\gamma,k}_{{\bbox{\delta}}_3} \;,
  \label{eq:def_tijk}
\end{equation}
the scaling Eq.~(\ref{eq:sca_tddd}) can be diagonalized and we obtain a set
of decoupled linear differential equations for the $\tilde
t^\gamma_{ijk}(\omega)$
\begin{equation}
  \omega
  { \partial \tilde t^\gamma_{ijk}(\omega)
     \over \partial \omega}
  =  (\lambda_i+\lambda_j-\lambda_k)\rho
      \tilde t^\gamma_{ijk}(\omega) \;,
  \label{eq:sca_tijk}
\end{equation}
whose solution is
\begin{equation}
  \tilde t^\gamma_{ijk}(\omega)=
   \tilde t^{\gamma(0)}_{ijk}
   \left( D \over \omega
    \right)^{\displaystyle(\lambda_i+\lambda_j-\lambda_k)\rho} \;.
  \label{eq:sol_tijk}
\end{equation}
Putting Eqs.~(\ref{eq:def_tddd}), (\ref{eq:def_tijk}) and
(\ref{eq:sol_tijk}) together, we get the fully factorized form
of the effective assisted hopping:
\begin{equation}
  \tilde
    t^\gamma(\bbox{\kappa}_1,\bbox{\kappa}_2,\bbox{\kappa}_3,\omega)
    = \sum_{ijk}
    \xi^{\gamma}_i(\bbox{\kappa}_1)
    \xi^{\gamma}_j(\bbox{\kappa}_2)
    \xi^{*\gamma}_k(\bbox{\kappa}_3)
   \left( D \over \omega \right)^{(\lambda_i+\lambda_j-\lambda_k)\rho}
   \tilde t^{\gamma(0)}_{ijk} \;,
  \label{eq:sol_tkkk}
\end{equation}
where we have introduced
\begin{equation}
  \xi^{\gamma}_j(\bbox{\kappa}) =
  \sum_{{\bbox{\delta}}}
  \phi^*_{{\bbox{\delta}}}(\bbox{\kappa})
  s^{\gamma,j}_{{\bbox{\delta}}}
  {\rm e}^{
      -i {\bf k}^{\vphantom{\dagger}}_F(\bbox{\kappa})\bbox{\delta} a/2} \;.
  \label{eq:def_xi}
\end{equation}
The exponents are of the order of $\rho U$, since in the
eigenvalue Eq.~(\ref{eq:eigenFU}) ${\sf F}$ is of the order of
unity, thus the magnitude of $\lambda$ is determined by
$U$.

The couplings $\tilde t^{\gamma(0)}_{ijk}$ can be determined from the initial
(unrenormalized) couplings (with $D$ being the bandwidth)
$\tilde
t^{\gamma(0)}(\bbox{\kappa}_1,\bbox{\kappa}_2,\bbox{\kappa}_3)=\tilde
t^{\gamma}(\bbox{\kappa}_1,\bbox{\kappa}_2,\bbox{\kappa}_3,\omega=D)$
[see Eqs.~(\ref{eq:asst_unren}) and (\ref{eq:sol_tkkk})],
 since in that case
\begin{equation}
\tilde t^{\gamma(0)}(\bbox{\kappa}_1,\bbox{\kappa}_2,\bbox{\kappa}_3,D)=
   \sum_{ijk}
    \xi^{\gamma}_i(\bbox{\kappa}_1)
    \xi^{\gamma}_j(\bbox{\kappa}_2)
    \xi^{*\gamma}_k(\bbox{\kappa}_3)
   \tilde t^{\gamma(0)}_{ijk}
  \label{eq:comptkkk} \;.
\end{equation}
After some algebraic manipulations we get
\begin{equation}
   \tilde t^{\gamma(0)}_{ijk} = \sum_{{\bbox{\delta}}}
      \tilde t^\gamma_{{\bbox{\delta}}}
    r^{\gamma,i}_{{\bbox{\delta}}} r^{\gamma,j}_{{\bbox{\delta}}}
    r^{*\gamma,k}_{{\bbox{\delta}}}
   \label{eq:calc_tijk0} \;.
\end{equation}
The latter equation tells us that $\tilde t^{\gamma(0)}_{ijk}$
is symmetrical in the first two indices (
$\tilde t^{\gamma(0)}_{ijk}=\tilde t^{\gamma(0)}_{jik}$) and as
a consequence [see Eq.~(\ref{eq:sol_tkkk})] the
$\tilde t^\gamma(\bbox{\kappa}_1,\bbox{\kappa}_2,\bbox{\kappa}_3,\omega)$
is also symmetrical in the first two $\bbox{\kappa}$ variables:
\begin{equation}
  \tilde
   t^\gamma(\bbox{\kappa}_1,\bbox{\kappa}_2,\bbox{\kappa}_3,\omega)
   = \tilde
   t^\gamma(\bbox{\kappa}_2,\bbox{\kappa}_1,\bbox{\kappa}_3,\omega)
   \;.
  \label{eq:t_sym}
\end{equation}

 For later use we mention, that the average of $\xi^*\xi$ can be
expressed as
\begin{equation}
  \bigl\langle
    \xi^{*\gamma}_j (\bbox{\kappa}) \xi^{\gamma}_{j'} (\bbox{\kappa})
  \bigr\rangle_{\bbox{\kappa}}
 = \sum_{{\bbox{\delta}}{\bbox{\delta}}'}
   s^{*\gamma,j}_{\bbox{\delta}}
   F_{{\bbox{\delta}}{\bbox{\delta}}'}
   s^{\gamma,j'}_{{\bbox{\delta}}'} \;.
   \label{eq:xixi}
\end{equation}

A similar approach was used by Dzyaloshinski \cite{Dzhalo87}
in case of the two dimensional nearly nested
electron gas. However, unlike
in our case, those scaling equations do not form a closed set.

\section{LIGHT PARTICLE SELF-ENERGY}

Unlike to usual renormalization procedure, where the self--energy
corrections become important only in the next to leading
logarithmic order, in our case we expect a large mass
renormalization for the light electrons as the logarithmic corrections in the
self--energy and in the generated electron--electron interaction occur on the
same level.

  The $l$-electron
self--energy diagrams are shown in Fig.~\ref{fig:self_ene}.  We
are going to evaluate them for those electrons which are close to
the Fermi level and we assume that $|\varepsilon_h|\ll \varepsilon_F$ or for
energies close to
$\varepsilon_h$. The latter case is of no importance for later
calculation and we are presenting it just for being interesting.

The heavy particle self--energy of
the lowest order contains three $l$-electron lines in the
intermediate state, thus it is nonlogarithmic except in one
dimension.

\subsection{Self--energy near the Fermi-level}

In the
logarithmic approximation the intermediate state with the two smallest
$l$-electron energy variables $\xi_1$ and $\xi_2$ must be picked up and then
both vertices must be fully renormalized, but their lower cutoff must be
replaced by $|\varepsilon_h|$. A typical integral
to be calculated for the self--energy with energy $\omega\ll\varepsilon_h$ is
\begin{eqnarray}
 \int_{\varepsilon_h}^Dd\xi_1 \int_{\varepsilon_h}^Dd\xi_2 &&
  \left(D \over \xi_1+\xi_2 \right)^{\alpha}
  { \pm \mbox{sign}(\varepsilon_h) \over
   \pm \mbox{sign} (\varepsilon_h) \omega - (\xi_1+\xi_2+|\varepsilon_h|)}=
   \nonumber\\
  && =-(\omega \mp \varepsilon_h) A (D / |\varepsilon_h| ) \;,
    \label{eq:e_int}
\end{eqnarray}
where the sign is appropriately chosen to give the correct
contributions of different diagrams: the upper sign stands for the
first two diagrams and lower sign for the last in Fig.~\ref{fig:self_ene}.
  In the actual calculation we pick up the most divergent term,
   so that the largest exponent is taken as $\alpha$.
The vertex contributions contains also $\omega$ variables, but its sign
is not well defined. It is easy to demonstrate that the appearance
of such an $\omega$ changes the right hand side of
Eq.~(\ref{eq:e_int}), by factors like $(1\pm\alpha)^{-1}$.  As
the logarithmic approximation is applied, therefore, the
corrections proportional to $U$ must be neglected. Furthermore,
the shift of chemical potential $\varepsilon_h A$ is also only an
approximate value, since the in this approximation it is
difficult to determine exactly \cite{NodeDo69}. The
coefficient $A$ can be calculated and
\begin{equation}
  A(D/|\varepsilon_h|)={1\over \alpha}
  \left[\left(D \over |\varepsilon_h|\right)^{\alpha}-1\right]
  \label{eq:A}
\end{equation}
is obtained.

The complete expression for the self--energy is
\begin{eqnarray}
  \Sigma_l(\bbox{\kappa},\omega) & = &
  \sum_{\gamma}
  \int_{{\rm Max}(\omega,\varepsilon_h)}^Dd\varepsilon \; \rho^2 \nonumber\\
  \times  && \left(
   \Bigl\langle
     \tilde
t^\gamma(\bbox{\kappa},\bbox{\kappa}',\bbox{\kappa}'';\varepsilon)
     {1\over\varepsilon}
     \tilde t^{\gamma
*}(\bbox{\kappa},\bbox{\kappa}',\bbox{\kappa}'';\varepsilon)
   \Bigr\rangle_{\bbox{\kappa}'\bbox{\kappa}''} \right.\nonumber\\
   && +
   \Bigl\langle
     \tilde
t^\gamma(\bbox{\kappa}',\bbox{\kappa},\bbox{\kappa}'';\varepsilon)
     {1\over\varepsilon}
     \tilde t^{\gamma
*}(\bbox{\kappa}',\bbox{\kappa},\bbox{\kappa}'';\varepsilon)
   \Bigr\rangle_{\bbox{\kappa}'\bbox{\kappa}''} \nonumber\\
   && -
   \left.
   \Bigl\langle
     \tilde
t^\gamma(\bbox{\kappa}',\bbox{\kappa}'',\bbox{\kappa};\varepsilon)
     {1\over\varepsilon}
     \tilde t^{\gamma
*}(\bbox{\kappa}',\bbox{\kappa}'',\bbox{\kappa};\varepsilon)
   \Bigr\rangle_{\bbox{\kappa}'\bbox{\kappa}''}
 \right)  \;,
 \label{eq:self_e}
\end{eqnarray}
where the different terms on the right hand side  comes from the three diagrams
in
Fig.~\ref{fig:self_ene}, respectively.
Performing the integration, we get
\begin{eqnarray}
  \Sigma_l(\bbox{\kappa};\omega) & = &
  -\omega[
      \chi_1(\bbox{\kappa})
     +\chi_2(\bbox{\kappa})
     +\chi_3(\bbox{\kappa})
   ]  \rho^2 A(D/\varepsilon_h) \nonumber \\
  & + & \varepsilon_h
    [ \chi_1(\bbox{\kappa})
     +\chi_2(\bbox{\kappa})
     -\chi_3(\bbox{\kappa})]
    \rho^2 A(D/\varepsilon_h) \label{eq:self_e_chi} \;,
    \label{eq:self_e_int}
\end{eqnarray}
where
\begin{equation}
  \left.
    \begin{array}{c}
      \chi_1(\bbox{\kappa}) \\
      \chi_2(\bbox{\kappa}) \\
      \chi_3(\bbox{\kappa}) \\
    \end{array}
  \right\}
  = \sum_\gamma
  \left\{
    \begin{array}{c}
      \bigl\langle
       \tilde t^\gamma(\bbox{\kappa},\bbox{\kappa}',\bbox{\kappa}'')
       \tilde t^{\gamma*}(\bbox{\kappa},\bbox{\kappa}',\bbox{\kappa}'')
      \bigr\rangle_{\bbox{\kappa}'\bbox{\kappa}''}
    \\
      \bigl\langle
       \tilde t^\gamma(\bbox{\kappa}',\bbox{\kappa},\bbox{\kappa}'')
       \tilde t^{\gamma*}(\bbox{\kappa}',\bbox{\kappa},\bbox{\kappa}'')
      \bigr\rangle_{\bbox{\kappa}'\bbox{\kappa}''}
     \\
       \bigl\langle
        \tilde t^\gamma(\bbox{\kappa}',\bbox{\kappa}'',\bbox{\kappa})
        \tilde t^{\gamma*}(\bbox{\kappa}',\bbox{\kappa}'',\bbox{\kappa})
       \bigr\rangle_{\bbox{\kappa}'\bbox{\kappa}''}
     \\
    \end{array}
  \right. \;,
  \label{eq:def_chi}
\end{equation}
and, furthermore, the most divergent term of $\tilde t^\gamma$
[see Eq.~(\ref{eq:sol_tkkk})] is used (the exponent $\alpha$ is the
largest of the eigenvalue combination $\lambda_i+\lambda_j-\lambda_k$),
which is a good approximation for most of the cases.

 The symmetry of $\tilde t^\gamma$ [see Eq.~(\ref{eq:t_sym})] implies
$\chi_1(\bbox{\kappa})=\chi_2(\bbox{\kappa})$,
furthermore, the average over the Fermi surface of every
$\chi_j(\bbox{\kappa})$ is equal and will be denoted by $\chi$.

Knowing the self--energy given by Eq.~(\ref{eq:self_e_chi}), the
renormalized one particle Green's function is obtained as
\begin{equation}
  G_l({\bf k},\omega)={
    Z_{\bbox{\kappa}}
    \over
    \omega-v^{\vphantom{\dagger}}_{F,{\rm ren}}(\bbox{\kappa})
      [k-k^{\vphantom{\dagger}}_F(\bbox{\kappa})]}
   \;,
  \label{eq:ren_G}
\end{equation}
where we have neglected the renormalization of the chemical
potential, which comes from the second term in
Eq.~(\ref{eq:self_e_int}) and it is hard to calculate
in leading logarithmic order approximation.
 The renormalization constant is defined as
\begin{eqnarray}
  Z^{-1}_{\bbox{\kappa}}
  & = & 1-
    \left.
      {\partial \mbox{Re} \Sigma_l(\omega,k) \over \partial \omega}
    \right|_{\omega=0} \nonumber\\
  & = & 1 + [2\chi_1(\bbox{\kappa})+\chi_3(\bbox{\kappa})]
  \rho^2
  A(D/|\varepsilon_h|)\;>\;1  \;, \label{eq:Z}
\end{eqnarray}
and the Fermi velocity is renormalized nearby the Fermi surface as
\begin{equation}
  v^{\vphantom{\dagger}}_{F,{\rm ren}}(\bbox{\kappa})=
   v^{\vphantom{\dagger}}_F(\bbox{\kappa})
   Z_{\bbox{\kappa}}\;<\;
   v^{\vphantom{\dagger}}_F(\bbox{\kappa}) \;,
  \label{eq:vF_ren}
\end{equation}
thus it is suppressed, leading to mass enhancement. The
dispersion curve is schematically plotted in
Fig.~\ref{fig:disp_curve}.  In the case of the large
renormalization described by $Z_{\bbox{\kappa}}\ll 1$ the large
modification of the dispersion curve is expected for
$|\omega|<|\varepsilon_h|$ and  the
renormalization gradually disappears as approaching larger
energies, $|\omega|>|\varepsilon_h|$.
Thus, the large enhancement of the density of states is restricted to a small
energy region
$|\omega|<|\varepsilon_h|Z_{\bbox{\kappa}}$.

The average strength of the mass enhancement can be given by
\begin{equation}
  \langle Z^{-1}_{\bbox{\kappa}} \rangle =
  1 + 3 \chi \rho^2 A(D/|\varepsilon_h|) \;.
  \label{eq:mass_inh}
\end{equation}

The mass renormalization can be very large, thus the scaling may result
in heavy fermionic behavior. The calculation presented are justified only in
the region $\alpha\ll1$, but there is no indication for nonanalytical
behavior, thus in order to get qualitative result the extrapolation
for intermediate coupling $\alpha<1/2$ is adequate. As it is known
from the X-ray absorption problem \cite{NodeDo69}, for large $U\rho$
that quantity must be replaced by $\delta/\pi$ where $\delta$ is the
phase shift $(\delta\le\pi/2)$.

In Eq.~(\ref{eq:ren_G}) there is also a shift of the energy
proportional to
$\chi_1(\bbox{\kappa})+\chi_2(\bbox{\kappa})-\chi_3(\bbox{\kappa})$ which
is sensitive on the direction $\bbox{\kappa}$.  That shift changes
the number of the electrons inside the Fermi surface, thus the
Fermi energy must be corrected in order of
$\rho^2\chi \varepsilon_h A(D/\varepsilon_h)Z_{\bbox{\kappa}}$ to keep the
number
of particles inside the Fermi surface unchanged and that is
associated also with the deformation of shape of the Fermi
surface.

Finally it must be mentioned, that even a large wave function
renormalization does not play an important role in the vertex equation
for $\tilde t$, as the Green's functions are taken between two points in
the real space not further apart than the lattice constant. In that
case the Fourier transform of the Green's function multiplied by a
slowly varying function like $\exp(ik_xa/2)$ must be integrated with
respect to the momentum. The result is almost independent of the
self--energy and the correction is $O({\mit\Sigma}/D)$.

\subsection{Self--energy near the singularity}

If the energy of the electron is large enough, than the
scattering of an electron
from the filled Fermi sea to the unoccupied heavy level becomes a
real process, and we are
faced with a problem very similar to that of the X--ray edge \cite{NodeDo69},
where a power--law behavior is observed in the absorption
function. In our case the role of the X--ray is taken over by the
 incoming light electron, and power--like behavior is expected
in the imaginary part of the self--energy and so in the spectral
functions. For simplicity we are going to consider the
$\varepsilon_h>0$ case here. Similar considerations are valid if
$\varepsilon_h<0$.

  It is easy to see, that in
Fig.~\ref{fig:self_ene} for $\omega\approx\varepsilon_h$ the first
diagram
$\Sigma_{l,1}(\omega,\bbox{\kappa})$ and second diagram
$\Sigma_{l,2}(\omega,\bbox{\kappa})$, while for $\omega\approx-\varepsilon_h$
the
third diagram $\Sigma_{l,3}(\omega,\bbox{\kappa})$ is singular.
In this case the integral in (\ref{eq:e_int}) is
\begin{eqnarray}
 \int_{|\omega-\varepsilon_h|}^D d\xi_1 \int_{|\omega-\varepsilon_h|}^Dd\xi_2
&&
  \left(D \over \xi_1+\xi_2 \right)^{\alpha}
  { \pm 1\over
       \pm\omega-(\xi_1+\xi_2+|\varepsilon_h|)}
  = \nonumber\\
  && =-(\omega \mp \varepsilon_h) A (D / |\omega\mp\varepsilon_h| )
    \bigl[1\pm i\alpha\pi\Theta(\pm\omega\varepsilon_h)\bigr] \;,
    \label{eq:e_intX}
\end{eqnarray}
where the $\Theta (x)$ is the step function and
it ensures the proper analytical behavior.
The signs are the same as they were in Eq.~(\ref{eq:e_int}).

For this special choice of energies the renormalization of the
assisted hopping [Eq.~(\ref{eq:sca_tk1k2k3})] should be considered
more carefully. It turns out that the energies of the internal
electron lines are such, that for $\Sigma_{l,1}(\omega,\bbox{\kappa})$
the first term in Eq.~(\ref{eq:sca_tddd}) is not singular, similarly
for $\Sigma_{l,2}(\omega,\bbox{\kappa})$ the second term and for the
$\Sigma_{l,3}(\omega,\bbox{\kappa})$ the third term
can be neglected, so that the exponent $\lambda_i+\lambda_j-\lambda_k$ is
reduced to the values $\lambda_j-\lambda_k$, $\lambda_i-\lambda_k$ and
$\lambda_i+\lambda_j$ respectively. The exponent $\alpha$ should
be associated with the largest exponent for each case
separately and will be denoted by $\alpha_1=\alpha_2$ and
$\alpha_3$, and $A_i$ is the function defined by Eq.~(\ref{eq:A}) with
$\alpha=\alpha_i$. Putting everything together,
the self--energy for energies near $\omega=\varepsilon_h$ is
\begin{equation}
  \Sigma_{l}(\bbox{\kappa};\omega) =
   -(\omega-\varepsilon_h)
    \bigl[\chi_1(\bbox{\kappa})+\chi_2(\bbox{\kappa})\bigr]
    \rho^2 A_1(D/|\omega-\varepsilon_h|)
    \bigl[1+i\alpha_1\pi\Theta(\omega-\varepsilon_h)\bigr]
    \label{eq:self_e_X1}
\end{equation}
and
\begin{equation}
  \Sigma_{l}(\bbox{\kappa};\omega) =
    -(\omega+\varepsilon_h)\chi_3(\bbox{\kappa})
    \rho^2 A_3(D/|\omega+\varepsilon_h|)
    \bigl[1-i\alpha_3\pi\Theta(-\omega-\varepsilon_h)\bigr]
        \label{eq:self_e_X3}
\end{equation}
for $\omega\approx-\varepsilon_h$. The $\chi(\bbox{\kappa})$-s are
the ones defined by Eq.~(\ref{eq:def_xi}), however the terms
to be averaged are not the same and, therefore, the
averages are no more equal.

The imaginary parts of the self--energies (\ref{eq:self_e_X1})
contribute to the spectral densities of the particles and
it emerges as soon as the energy of the electron is larger then $\approx
\varepsilon_h$, while the self--energy (\ref{eq:self_e_X3}) gives similar
contribution for the spectral density of the holes having energy
below the threshold $-\varepsilon_h$.

\section{GENERATED INTERACTION BETWEEN THE LIGHT ELECTRONS}

To address  the possibility of superconductivity in light band
in these models, we need to calculate the generated
interaction in this band for parallel-- and anti--parallel spin.
The typical diagrams are shown in Fig.~\ref{fig:gen_vertex}. In
the logarithmic approximation the intermediate state with one $h$--electron and
the $l$--electron with the smallest energy must be picked up and
in vertices that energy must must be used as the lower cutoff.

The schematic form of the generated interaction is
\begin{equation}
  \int_\omega^Dt(\varepsilon)
  {1\over-\varepsilon}t(\varepsilon)d\varepsilon
  \sim-t^2(\varepsilon)
\end{equation}
without numerical factors containing the strength of the Coulomb
interaction.

The calculation is
straightforward in the time-ordered diagram technique. For
the interaction in the anti-parallel spin channel
the contribution of the four diagrams in Fig.~\ref{fig:gen_vertex}(a) is
\begin{eqnarray}
&&V^\bot(\bbox{\kappa}_1,\bbox{\kappa}_2,\bbox{\kappa}_3,\bbox{\kappa}_4)
 =  - \rho \sum_{\gamma}
  \int_{{\rm Max}(\omega,\varepsilon_h)}^Dd\varepsilon \nonumber\\
  \times  \biggl(
   &&\bigl\langle
     \tilde
t^\gamma(\bbox{\kappa}_1,\bbox{\kappa}_2,\bbox{\kappa};\varepsilon)
     {1\over\varepsilon}
     \tilde
  t^{\gamma*}(\bbox{\kappa}_4,\bbox{\kappa}_3,\bbox{\kappa};\varepsilon)
   \bigr\rangle_{\bbox{\kappa}}
   +
    \bigl\langle
     \tilde
t^\gamma(\bbox{\kappa}_2,\bbox{\kappa}_1,\bbox{\kappa};\varepsilon)
     {1\over\varepsilon}
\tilde t^{\gamma*}(\bbox{\kappa}_3,\bbox{\kappa}_4,\bbox{\kappa};\varepsilon)
   \bigr\rangle_{\bbox{\kappa}}
  \nonumber\\
  - &&
   \bigl\langle
     \tilde
t^\gamma(\bbox{\kappa}_1,\bbox{\kappa},\bbox{\kappa}_3;\varepsilon)
     {1\over\varepsilon}
     \tilde
    t^{\gamma *}(\bbox{\kappa}_4,\bbox{\kappa},\bbox{\kappa}_2;\varepsilon)
   \bigr\rangle_{\bbox{\kappa}}
  -
   \bigl\langle
     \tilde
t^\gamma(\bbox{\kappa}_2,\bbox{\kappa},\bbox{\kappa}_4;\varepsilon)
     {1\over\varepsilon}
     \tilde
 t^{\gamma *}(\bbox{\kappa}_3,\bbox{\kappa},\bbox{\kappa}_1;\varepsilon)
   \bigr\rangle_{\bbox{\kappa}}
 \biggr)  \;.
 \label{eq:Vap}
\end{eqnarray}

 The evaluation of diagrams shown in Fig.~\ref{fig:gen_vertex}(b)
gives in the parallel spin channel
\begin{eqnarray}
 && V^\Vert(\bbox{\kappa}_1,\bbox{\kappa}_2,\bbox{\kappa}_3,\bbox{\kappa}_4)
   = - {\rho\over 2} \sum_{\gamma}
  \int_{{\rm Max}(\omega,\varepsilon_h)}^Dd\varepsilon \nonumber\\
  \times && \biggl(
   \bigl\langle
     \tilde
  t^\gamma(\bbox{\kappa},\bbox{\kappa}_1,\bbox{\kappa}_4;\varepsilon)
     {1\over\varepsilon}
     \tilde
    t^{\gamma *}(\bbox{\kappa},\bbox{\kappa}_3,\bbox{\kappa}_2;\varepsilon)
   \bigr\rangle_{\bbox{\kappa}}
  +
   \bigl\langle
     \tilde
t^\gamma(\bbox{\kappa},\bbox{\kappa}_2,\bbox{\kappa}_3;\varepsilon)
     {1\over\varepsilon}
     \tilde
  t^{\gamma*}(\bbox{\kappa},\bbox{\kappa}_4,\bbox{\kappa}_1;\varepsilon)
   \bigr\rangle_{\bbox{\kappa}}
  \nonumber\\  - &&
   \bigl\langle
     \tilde
t^\gamma(\bbox{\kappa},\bbox{\kappa}_1,\bbox{\kappa}_3;\varepsilon)
     {1\over\varepsilon}
     \tilde
 t^{\gamma *}(\bbox{\kappa},\bbox{\kappa}_4,\bbox{\kappa}_2;\varepsilon)
   \bigr\rangle_{\bbox{\kappa}}
  -
   \bigl\langle
     \tilde
t^\gamma(\bbox{\kappa},\bbox{\kappa}_2,\bbox{\kappa}_4;\varepsilon)
     {1\over\varepsilon}
     \tilde
  t^{\gamma *}(\bbox{\kappa},\bbox{\kappa}_3,\bbox{\kappa}_1;\varepsilon)
   \bigr\rangle_{\bbox{\kappa}}
 \biggr) \;,
 \label{eq:Vp}
\end{eqnarray}
where we have antisymmetrized the interaction.
These expressions can be evaluated by using
Eq.~(\ref{eq:sol_tkkk}) for the scattering amplitude. The
induced interaction is constant in the energy range
$|\omega|<Z|\varepsilon_h|$ and drops rapidly for higher energies.

The interaction can be split into two terms: the interaction $V^S$
between electrons forming a singlet and $V^T$ between electrons
in a triplet state.  In the singlet state the spin part of
the wave function is antisymmetrized, so the $V^S$ is symmetrical
in the $\bbox{\kappa}_1$ and $\bbox{\kappa}_2$ variables
\begin{equation}
   V^S(\bbox{\kappa}_1,\bbox{\kappa}_2,\bbox{\kappa}_3,\bbox{\kappa}_4)
 = V^S(\bbox{\kappa}_2,\bbox{\kappa}_1,\bbox{\kappa}_3,\bbox{\kappa}_4)
\;.
\end{equation}
For the triplet case the symmetrycal spin wavefunction
implies antisymmetrycal property of the $V^T$, so
\begin{equation}
   V^T(\bbox{\kappa}_1,\bbox{\kappa}_2,\bbox{\kappa}_3,\bbox{\kappa}_4)
= -V^T(\bbox{\kappa}_2,\bbox{\kappa}_1,\bbox{\kappa}_3,\bbox{\kappa}_4)\;,
\end{equation}
 and similar holds for the last two $\bbox{\kappa}$ variables.
  Furthermore, it is evident that for both interactions
\begin{equation}
V^{S,T}(\bbox{\kappa}_1,\bbox{\kappa}_2,\bbox{\kappa}_3,\bbox{\kappa}_4) =
V^{S,T}(\bbox{\kappa}_2,\bbox{\kappa}_1,\bbox{\kappa}_4,\bbox{\kappa}_3)\;.
\end{equation}

 The  $V^\Vert$ is by definition
the interaction between electrons in triplet state. To
$V^\bot$, however, both the singlet and triplet states
contributes. We can split them by using the symmetry
of the $V^S$ and $V^T$:
\begin{equation}
V^S(\bbox{\kappa}_1,\bbox{\kappa}_2,\bbox{\kappa}_3,\bbox{\kappa}_4)
=\frac{1}{2}
  \bigl[
    V^\bot(\bbox{\kappa}_1,\bbox{\kappa}_2,\bbox{\kappa}_3,\bbox{\kappa}_4)
  + V^\bot(\bbox{\kappa}_2,\bbox{\kappa}_1,\bbox{\kappa}_3,\bbox{\kappa}_4)
  \bigr] \;,
\end{equation}
and
\begin{equation}
  V^T(\bbox{\kappa}_1,\bbox{\kappa}_2,\bbox{\kappa}_3,\bbox{\kappa}_4)
  =\frac{1}{2}
  \bigl[
    V^\bot(\bbox{\kappa}_1,\bbox{\kappa}_2,\bbox{\kappa}_3,\bbox{\kappa}_4)
  - V^\bot(\bbox{\kappa}_2,\bbox{\kappa}_1,\bbox{\kappa}_3,\bbox{\kappa}_4)
  \bigr] \;.
\end{equation}
The latter must be exactly
$V^\Vert(\bbox{\kappa}_1,\bbox{\kappa}_2,\bbox{\kappa}_3,\bbox{\kappa}_4)$,
as it
can be seen if we take into account the symmetries of $\tilde t^\gamma$
[Eq.~(\ref{eq:t_sym})]. This separation is possible because of
the absence of the spin-orbit coupling in this model.

Performing the energy integral, we get for the interaction in
the singlet channel
\begin{eqnarray}
  &&V^S(\bbox{\kappa}_1,\bbox{\kappa}_2,\bbox{\kappa}_3,\bbox{\kappa}_4)
  =  - {\rho\over 2}  \sum_{\gamma} A(D/\varepsilon_h)
   \biggl(
   4 \bigl\langle
     \tilde t^\gamma(\bbox{\kappa}_1,\bbox{\kappa}_2,\bbox{\kappa})
     \tilde t^{\gamma *}(\bbox{\kappa}_4,\bbox{\kappa}_3,\bbox{\kappa})
   \bigr\rangle_{\bbox{\kappa}} \nonumber\\
   &&-
   \bigl\langle
     \tilde t^\gamma(\bbox{\kappa}_1,\bbox{\kappa},\bbox{\kappa}_3)
     \tilde t^{\gamma *}(\bbox{\kappa}_4,\bbox{\kappa},\bbox{\kappa}_2)
   \bigr\rangle_{\bbox{\kappa}}
   -
   \bigl\langle
     \tilde t^\gamma(\bbox{\kappa}_2,\bbox{\kappa},\bbox{\kappa}_4)
     \tilde t^{\gamma *}(\bbox{\kappa}_3,\bbox{\kappa},\bbox{\kappa}_1)
   \bigr\rangle_{\bbox{\kappa}}
   \nonumber\\
   &&-
   \bigl\langle
     \tilde t^\gamma(\bbox{\kappa}_2,\bbox{\kappa},\bbox{\kappa}_3)
     \tilde t^{\gamma *}(\bbox{\kappa}_4,\bbox{\kappa},\bbox{\kappa}_1)
   \bigr\rangle_{\bbox{\kappa}}
   -
   \bigl\langle
     \tilde t^\gamma(\bbox{\kappa}_1,\bbox{\kappa},\bbox{\kappa}_4)
     \tilde t^{\gamma *}(\bbox{\kappa}_3,\bbox{\kappa},\bbox{\kappa}_2)
   \bigr\rangle_{\bbox{\kappa}}
 \biggr)
 \label{eq:VS}
\end{eqnarray}
and for the triplet channel the result is identical with Eq.~(\ref{eq:Vp})
where the energy integral must be replaced by $A(D/\varepsilon_h)$.

We can see that if the assisted hopping $\tilde t^\gamma$ is
structureless, than the generated interaction disappears both
in singlet and triplet
channel. However, a small $\bbox{\kappa}$ dependence can lead to
finite interactions
which may be  enhanced due to large $A(D/|\varepsilon_h|)$.

\section{SUPERCONDUCTING TRANSITION TEMPERATURE}

As it has been mentioned in Sec.II the induced interaction
between the light particles (Fig.~\ref{fig:gen_vertex}) and the
self--energy (Fig.~\ref{fig:self_ene}) are connected by identities
similar to the Ward
identities. The Ward identities are valid
only in the electron-hole channel with zero total spin and not
in the Cooper channel.
These relations are due to the fact, that both occurs first in
the second order of the perturbation theory on $\tilde t$. That
is in contrast to most of the other logarithmic theories, where
the vertex correction exists already in the first order. Thus in
the present case, the induced vertex and the self--energy must be
treated simultaneously in determinations of the superconducting
order parameter $\Delta$ and of the transition temperature, which
are based on the selfconsistent
equation shown by diagrams in Fig.~\ref{fig:gap_eq}.

The order parameter is defined as $\Delta_{\sigma\sigma'}({\bf k})=
\langle c_{{\bf k},\sigma}c_{-{\bf k},\sigma'}\rangle$ ,
where $\sigma$ and $\sigma'$ stand for spins. Separating the
spin wave function,  $\Delta$
is either even or odd function of $\bf k$, depending whether we are
looking at singlet ($\Delta^S$) or triplet ($\Delta^T$)
Cooper-pairs. Using the interaction $V^{S(T)}$, the
self-consistent equation for $\Delta^{S(T)}$ is
\begin{eqnarray}
  \Delta^{S(T)}({\bf k})
   &=& - T_c {1\over N}\sum_{\bf k'}
  \sum_{\omega_m} V^{S(T)}
  ({\bf k},-{\bf k},-{\bf k}',{\bf k}') \nonumber\\
  &\times& G({\bf k'};\omega_m)G(-{\bf k'};\omega_n-\omega_m)
  \Delta^{S(T)}({\bf k'}) \label{eq:Tc_selfcons}\;,
\end{eqnarray}
with $T_c$ being the transition temperature and
$\omega_m=(2m+1)\pi T$ are the Matsubara frequencies.
After summing over the internal energy, we arrive at
\begin{equation}
  \Delta^{S(T)}(\bbox{\kappa})
   = - T_c \rho \sum_{\omega_m<\omega_c} {\pi\over|\omega_m|}
   \bigl\langle
     V^{S(T)}(\bbox{\kappa},-\bbox{\kappa},-\bbox{\kappa}',\bbox{\kappa}')
     Z(\bbox{\kappa}')
     \Delta^{S(T)}(\bbox{\kappa}')
   \bigr\rangle_{\bbox{\kappa}'} \;,
  \label{eq:Tc_selfcons1}
\end{equation}
where we introduced the frequency cutoff $\omega_c$ as we have
replaced the $\omega_m$ dependent quantities by their low
frequency values, assuming that $T_c$ is much smaller than the
characteristic energy $\varepsilon_h$. In our case
the  $\omega_c$ is determined by the
fact, that the contributions for the integral over the internal energy
comes from energies smaller than
$Z|\varepsilon_h|$, as it is determined by the linear part of the dispersion
curve where the density of states is enhanced
(Fig.~\ref{fig:disp_curve}),
 so the summation over the frequencies yields
\begin{eqnarray}
  \Delta^{S(T)}(\bbox{\kappa})
   &=& - \ln \left( 1.13 |\varepsilon_h|Z \over T_c \right)
   \rho
   \bigl\langle
     V^{S(T)}(\bbox{\kappa},-\bbox{\kappa},-\bbox{\kappa}',\bbox{\kappa}')
     Z(\bbox{\kappa}')
     \Delta^{S(T)}(\bbox{\kappa}')
   \bigr\rangle_{\bbox{\kappa}'} \;.
   \label{eq:Tc_selfcons_int}
\end{eqnarray}

In other words, in the selfconsistent equation for the order
parameter each interaction
$V$ is associated with two Greens's function. Only the
contribution of electrons with energy $\omega<\varepsilon_h$ is kept as the
vertex drops rapidly for larger energy. For the dominating low energy
electron the strength of the poles is $Z_{\bbox{\kappa}}$, and its
energy $|\omega|<\varepsilon_hZ_{\bbox{\kappa}}$ (see
Fig.~\ref{fig:disp_curve}). Furthermore, the
density of states $\rho$ for such electrons is also enhanced by
$Z_{\bbox{\kappa}}^{-1}$ [see Eq.~(\ref{eq:vF_ren})] .

In the singlet channel, where it can be assumed that the
$\bbox{\kappa}$--dependence of the superconducting order parameter
$\Delta(\bbox{\kappa})$ is weak,
the approximate strength of the dimensionless
effective coupling is obtained by
evaluating Eq.~(\ref{eq:Tc_selfcons_int}) so that
\begin{eqnarray}
  g^{(S)}_{\rm eff} &\sim& \rho
   \bigl\langle
     V^{S}(\bbox{\kappa},-\bbox{\kappa},-\bbox{\kappa}',\bbox{\kappa}')
   \bigr\rangle_{\bbox{\kappa},\bbox{\kappa}'}
   \bigl\langle
     Z(\bbox{\kappa}')
   \bigr\rangle_{\bbox{\kappa}'} \nonumber\\
    &\sim& \langle Z^{-1} \rangle^{-1}\langle V^S\rangle\rho \;,
    \label{eq:gseff}
\end{eqnarray}
where in the last approximation $Z_{\bbox{\kappa}}$ is replaced by
its average over the Fermi surface [see
Eq.~(\ref{eq:mass_inh})]. If the $\bbox{\kappa}$--dependence of the
$V$ and $Z$ is not large, then the average of $V^S$ is of the
order of $\rho\chi A$. Inserting the calculated value of
$Z_{\bbox{\kappa}}$ given by Eq.~(\ref{eq:mass_inh}), we can see
that the dimensionless coupling is
\begin{equation}
  g^{(S)}_{\rm eff} = q
  { \chi \rho^2  A(D/\varepsilon_h)
   \over
   1+3\chi \rho^2 A(D/\varepsilon_h)} \;,
   \label{eq:gSeff}
\end{equation}
where $q$ is of the order of unity and depends on the details on the
$\bbox{\kappa}$--dependencies, which is approximated in Eq.~(\ref{eq:gseff}).
If the mass renormalization is large,
$1+3\chi \rho^2 A(D/\varepsilon_h)\gg 1$,
 then the denominator is important and $g^{(S)}_{\rm eff}$ saturates.

 In this way in the BCS theory the transition temperature is
\begin{equation}
  T_c=|\varepsilon_h|\langle Z_{\bbox{\kappa}}\rangle \,
  {\rm e}^{\scriptstyle -{1/ g^{(S)}_{\rm eff}}}
\end{equation}
which can be expressed as a function of the averaged mass enhancement $\langle
Z_{\bbox{\kappa}}^{-1} \rangle$ given by Eq.~(\ref{eq:mass_inh}) and of the
parameter $q$. That function is shown in Fig.~\ref{fig:tc}. The transition
temperature $T_c$ is suppressed by the decreasing
$\langle Z_{\bbox{\kappa}}\rangle$ for strong
renormalization. The critical temperature is the largest in an intermediate
region of moderate renormalization where the mass enhancement
$\langle Z_{\bbox{\kappa}}^{-1} \rangle \sim 2-5$. As we have seen, in
general it is essential to include the renormalization of
the Fermi-velocity due to self--energy to get a consistent
treatment of the superconductivity.

\section{The 1D model}

To demonstrate the procedure described in the previous sections, let us
apply it to a 1-dimensional case introduced by one of the
authors \cite{Za89PR}.
In this model the light orbitals are on every site $n$ and the
heavy orbitals are on
every second site, as shown in Fig.~\ref{fig:1Dmod}. The
degeneracy of
the heavy orbitals is
neglected thus the index $\gamma$ will be dropped.
The hopping to heavy orbitals is
$t^\gamma_{h,\delta} = t_h $ and the hoppings between the
neighboring light
orbitals are $t$, the Coulomb repulsion
between the light and heavy orbitals is
\begin{equation}
  U_{\delta} =
    \left\{
      \begin{array}{ll}
        U  \;,\; &\mbox{if}\;\; \delta=0 \\
        U' \;,\; &\mbox{if}\;\; \delta=\pm 1 \quad , \\
      \end{array}
    \right.
\end{equation}
where the on--site and nearest neighbor interaction is kept.
Finally, the assisted hopping is
\begin{equation}
  \tilde t_{\delta} =
    \left\{
      \begin{array}{ll}
         0  \;,\;
          &\mbox{if}\;\; \delta=0 \\
         \tilde t \;,\;
          &\mbox{if}\;\; \delta=\pm 1 \quad . \\
      \end{array}
    \right.
  \label{eq:t_ass_1d_even}
\end{equation}
in case of light $s$ and heavy $d$ orbitals, where the wave functions have the
same sign in the regions of the overlap.
In the previous work \cite{Za89PR} $U'=0$ was taken.

The dispersion relation of the fermions in the light band, shown in
 Fig.~\ref{fig:1Ddisprel}, is given by
\begin{equation}
  \omega_\pm(k)= {\Delta\varepsilon \over 2} \pm
 \sqrt{
   \left(\Delta\varepsilon \over 2\right)^2 + 4t^2\cos^2 {k a \over 2}
 }\;
\end{equation}
in the tight binding approximation, where the $+(-)$ stands for
upper(lower) band, the Brillouin zone is $-\pi/a<k<\pi/a$, and
$\Delta \varepsilon=\varepsilon_\delta-\varepsilon_0$ is the relative shift of
the atomic levels.

As a next step, we
have to determine the amplitude $\phi(\bbox{\kappa})$ for $\delta=\pm 1,0$
[see Eq.~(\ref{eq:a=phi_d})]. In the
case of even $t_{\delta\delta'}$ the $\phi$-s are not $\kappa$ dependent
and  $\phi_{+1}=\phi_{-1}$. Since the probability is normalized,
$\phi_{\pm 1}^2+\phi_0^2=1$, we can parametrize $\phi$-s by
$\phi_{+1}=\phi_{-1}=\cos\varphi$ and $\phi_0=\sin\varphi$. Here
the subscript $+1$ ($-1$) stands for the right (left)
neighbors for the heavy orbital. If the energy levels of the
light orbitals are the same ($\varepsilon_\delta=0$), then all
the $\phi$-s are equal to $1/\sqrt{2}$ ($\varphi=\pi/4$). In the
general case the
$\varphi$ can be estimated as
\begin{equation}
  \tan \varphi= {\Delta\varepsilon\over 2 t \cos \vartheta} +
  \sqrt{ \left( \Delta\varepsilon \over 2 t \cos
     \vartheta\right)^2+1} \;,
\end{equation}
 where $\vartheta=k_Fa/2$ is introduced in accordance with
Ref.~\onlinecite{Za89PR}. It depends on the filling $\nu$, $\vartheta=\pi
\nu/2$ for the lower band ($\nu<1$) and $\vartheta=\pi (1-\nu/2)$ for
the upper band ($\nu>1$).

Specially, in one dimension, Eq.~(\ref{eq:sum2int}) reduces to
\begin{equation}
  {1 \over N} \sum_k = a \int{ d {\bf k} \over 2\pi}
  \quad\rightarrow\quad
  {a\over 2\pi}
  {1 \over v^{\vphantom{\dagger}}_F}
  \sum_{\bbox{\kappa} = \pm 1}
  \int^D_{-D}d\varepsilon \;,
  \label{eq:sum2int1d}
\end{equation}
where $\bbox{\kappa}=1$ and $-1$ denotes right and left moving
electrons, with $k^{\vphantom{\dagger}}_F(1)=k^{\vphantom{\dagger}}_F$ and
$k^{\vphantom{\dagger}}_F(-1)=-k^{\vphantom{\dagger}}_F$, similarly for the
velocities
$v^{\vphantom{\dagger}}_F(\pm 1)=\pm v^{\vphantom{\dagger}}_F$.  The form
factor appearing
in Eq.~(\ref{eq:expFS}) is also simplified significantly, since
there are only four combinations $\exp(\pm i \vartheta)$ and
$\exp(\pm i 3\vartheta)$.

 The density of states $\rho$
[Eqs.~(\ref{eq:rho})] is the usual one for the tight binding
approximation:
\begin{equation}
  \rho = {1\over 2\pi} {2 a \over v_F} \;,
\end{equation}
and the calculation of the amplitudes $F_{\delta\delta'}$
[Eq.~(\ref{eq:Fdd})] gives
\begin{eqnarray}
  F_{00} & = & \sin^2\varphi \nonumber\\
  F_{++} & = & F_{--}  =  \cos^2\varphi \nonumber\\
  F_{0+} & = & F_{0-} =  \cos \vartheta \cos\varphi\sin\varphi \nonumber\\
  F_{+-} & = & \cos 2\vartheta \cos^2\varphi \;.
\end{eqnarray}
The {\sf UF} matrix is then
\begin{equation}
{\sf U F} = \left(
        \begin{array}{ccc}
            U' \cos^2\varphi &
            U'\cos \vartheta \cos\varphi\sin\varphi &
            U' \cos 2\vartheta \cos^2\varphi\\
            U\cos \vartheta \cos\varphi\sin\varphi &
            U \sin^2\varphi &
            U\cos \vartheta \cos\varphi\sin\varphi  \\
            U' \cos 2\vartheta \cos^2\varphi &
            U'\cos \vartheta \cos\varphi\sin\varphi&
            U' \cos^2\varphi \\
        \end{array}
       \right) \;,
\end{equation}
The matrix ${\sf UF}$  is not symmetric, thus the left and right
eigenvectors are different, and one obtains for right and left
eigenvectors
\begin{equation}
 {\bf s}^{(1)} =
   \left[
     \begin{array}{c}
       U'\cos\varphi\cos\vartheta \\
       U \sin\varphi\\
       U' \cos\varphi\cos\vartheta \\
     \end{array}
   \right]
 \quad
 {\bf s}^{(2)} =
   \left[
     \begin{array}{c}
       \sin \varphi\\
       -2 \cos\varphi \cos\vartheta \\
       \sin\varphi \\
     \end{array}
   \right]
 \quad
 {\bf s}^{(3)} =
  \left[
     \begin{array}{c}
       1 \\
       0 \\
      -1 \\
     \end{array}
   \right]
\end{equation}
and
\begin{eqnarray}
 {\bf r}^{(1)} & = & {1\over \lambda_1}
   ( \cos\varphi\cos \vartheta \:,\; \sin\varphi \:,\;
     \cos\varphi\cos \vartheta ) \nonumber\\
 {\bf r}^{(2)} & = & {1\over 2\lambda_1}
    ( U\sin\varphi\:,\; -2U'\cos\varphi\cos\vartheta \:,\;  U \sin\varphi )
     \nonumber\\
 {\bf r}^{(3)} & = & ( 1/2 \:,\; 0 \:,\; -1/2 ) \;,
\end{eqnarray}
with eigenvalues $\lambda_1=U\sin^2\varphi+2U'\cos^2\varphi \cos^2\vartheta$,
$\lambda_2=0$ and $\lambda_3=2 U'\cos^2\varphi \sin^2\vartheta$.

The functions $\xi$  [see Eq.~(\ref{eq:def_xi})] are
\begin{eqnarray}
 \xi_1(\kappa) & = & \lambda_1 \nonumber\\
 \xi_2(\kappa) & = & 0 \nonumber\\
 \xi_3(\kappa) & = & - i 2 \sin \kappa \vartheta \cos \varphi \;.
\end{eqnarray}
The nonvanishing averages are
\begin{eqnarray}
 \langle|\xi_1(\kappa)|^2\rangle  & = & \lambda_1^2 \nonumber\\
 \langle|\xi_3(\kappa)|^2\rangle  & = & 4 \cos^2\varphi \sin^2 \vartheta
 \;.
\end{eqnarray}

{}From Eq.~(\ref{eq:calc_tijk0}), we get
\begin{eqnarray}
\tilde t^{(0)}_{111} & = & 2 \tilde t
   {1\over \lambda_1^3}\cos^3 \vartheta \cos^3\varphi \nonumber\\
\tilde t^{(0)}_{133} & = & \tilde t^{(0)}_{313} = \tilde t^{(0)}_{331} =
  {\tilde t \over 2} {1\over \lambda_1 }\cos \vartheta \cos \varphi  \;,
\end{eqnarray}
so that for the effective assisted hopping we get
\begin{eqnarray}
 \tilde t(\kappa_1,\kappa_2,\kappa_3;\omega)  = 2
  \tilde t \cos^3\varphi
    &&\left\{
       \bigl[ \cos^3 \vartheta + \cos \vartheta
          (\sin \kappa_1 \vartheta+\sin \kappa_2 \vartheta )
           \sin \kappa_3 \vartheta
       \bigr]  \bigl( D / \omega \bigr)^{\rho \lambda_1}
    \right. \nonumber \\
    && -  \left.
        \cos \vartheta \sin \kappa_1 \vartheta \sin \kappa_2 \vartheta
         \bigl( D / \omega \bigr)^{\rho (2\lambda_3-\lambda_1)}
       \right\} \;.
\end{eqnarray}
Since in the assisted hopping only the transitions of an
electron from one of the
neighboring sites to the heavy orbital play role, the amplitude
$\cos^3\varphi$ is
easy to understand being associated with the three light electron lines.
Furthermore, we can see that the Coulomb interactions
appear in the exponent only as the combinations $U' \cos^2\varphi$ and
$U\sin^2\varphi$, what means that the Coulomb
repulsions enters only as effective repulsion normalized by the
single site fermion densities.

As $\kappa$ can have the values $\pm 1$, the assisted hopping can
be given by the four amplitudes  $\tilde t_1$, $\tilde t_2$, $\tilde t_3$ and
$\tilde t_4$, introduced in Ref.~\onlinecite{Za89PR}. They can be expressed by
$
\tilde t(\kappa_1,\kappa_2,\kappa_3;\omega)$ as
\begin{eqnarray}
  \tilde t_1 &=& \tilde t(1,-1,1;\omega) =\tilde t^*(-1,1,-1;\omega)
\nonumber\\
  \tilde t_2 &=& \tilde t(1,-1,-1;\omega) =\tilde t^*(-1,1,1;\omega)
\nonumber\\
  \tilde t_3 &=& \tilde t(1,1,-1;\omega) =\tilde t^*(-1,-1,1;\omega)
\nonumber\\
  \tilde t_4 &=& \tilde t(1,1,1;\omega) =\tilde t^*(-1,-1,-1;\omega)
\end{eqnarray}
so that
\begin{eqnarray}
 \tilde t_1 &=& \tilde t_2 = 2 \tilde t \cos^3\varphi
   \left[
     \cos^3 \vartheta \bigl( D / \omega \bigr)^{\rho \lambda_1}
   + \cos \vartheta \sin^2 \vartheta
     \bigl( D / \omega \bigr)^{\rho(2\lambda_3-\lambda_1)}
   \right] \nonumber \\
 \tilde t_3 & =& 2
   \tilde t \cos^3\varphi
    \left[
       \bigl( \cos^3 \vartheta - 2 \cos \vartheta \sin^2 \vartheta
       \bigr)  \bigl( D / \omega \bigr)^{\rho\lambda_1}
    - \cos \vartheta \sin^2 \vartheta
         \bigl( D / \bigr)^{\rho(2\lambda_3-\lambda_1)} \right]
   \nonumber\\
 \tilde t_4 &=& 2
   \tilde t \cos^3\varphi
    \left[
       \bigl( \cos^3 \vartheta + 2 \cos \vartheta \sin^2 \vartheta
       \bigr)  \bigl( D / \omega \bigr)^{\rho\lambda_1}
    - \cos \vartheta \sin^2 \vartheta
         \bigl( D /\omega \bigr)^{\rho(2\lambda_3-\lambda_1)}
    \right] \;.
  \label{eq:1Devent}
\end{eqnarray}
These equation are exactly the solutions of the Eq.~(7) in
Ref.~\onlinecite{Za89PR} with initial couplings
$\tilde t_1^{(0)}=\tilde t_2^{(0)}=\tilde t_4^{(0)}=\tilde t
\cos^3\varphi\cos\vartheta$
, $\tilde t_3^{(0)}= \tilde t\cos^3\varphi \cos3\vartheta$, $U'=0$ and
$\varphi=\pi/4$.

As a next step we calculate the generated interactions between
the fermions in the light band. We are using the notations
common in the theory of the one--dimensional Fermi gas. For a
review, we refer to the paper of S\'olyom \cite{So79}.

 In the antiparallel channel the forward scattering,
denoted by $g_2$, is
\begin{equation}
  g_2=V^\bot(1,-1,-1,1) = 2|\tilde t_1|^2-2|\tilde t_3|^2 \;.
\end{equation}
 Replacing the assisted hopping by its most divergent part,
determined by the largest exponent, we get
\begin{equation}
 g_2 = (2\tilde t\cos^3\varphi)^2
  \left\{ \begin{array}{ll}
      2\sin^2 2\vartheta \cos 2 \vartheta
      (D/\varepsilon_h)^{2\lambda_1} \;,\;
          &\mbox{if}\;\; \lambda_1>\lambda_3 \\
      0\;,\;
           &\mbox{if}\;\; \lambda_1<\lambda_3 \;. \\
  \end{array} \right.
\end{equation}
If $\lambda_1>\lambda_3$, than $g_2$ is positive for small
fillings and became negative for larger fillings.

For the
backward scattering, where the momentum transfer is large
($2k_F$), we get
\begin{eqnarray}
  g_{1\bot} &=& V^\bot(1,-1,1,-1) =
    2 \tilde t_1 ( \tilde t_2- \tilde t_4)
   +2 \tilde t^*_1 ( \tilde t^*_2- \tilde t^*_4)
  \;,\nonumber\\
  g_{1\Vert} &=& V^\Vert(1,-1,1,-1)-V^\Vert(1,-1,-1,1)+g_2
   = 4 |\tilde t_1|^2 - 2 \tilde t^*_2 \tilde t_4
    -2 \tilde t_2 \tilde t^*_4
   \;.
\end{eqnarray}
 Since $t_1=t_2^*$, the $g_{1\bot}= g_{1\Vert}=g_1$ holds, which means
that the model is isotropic, and
\begin{equation}
 g_1 = (2\tilde t\cos^3\varphi)^2 \left\{ \begin{array}{ll}
       -\sin^2 2\vartheta (1+\cos 2\vartheta)
        (D/\varepsilon_h)^{2\lambda_1} \;,\;
        &\mbox{if} \;\; \lambda_1>\lambda_3 \\
        \sin^2 2\vartheta (1-\cos 2\vartheta)
        (D/\varepsilon_h)^{4\lambda_3-2\lambda_1} \;,\;
        &\mbox{if}\;\; \lambda_1<\lambda_3 \quad . \\
  \end{array} \right.
\end{equation}
They are presented in Fig.~\ref{fig:geff1D}.

 In one dimension the values of these couplings determine the
nature of the ground state \cite{So79}. We analyse the
$\lambda_1>\lambda_3$ and the $\lambda_3>\lambda_1$ region
separately. In the first case the $g_1$ is negative, and
depending whether the sign of the combination $g_1-2g_2$ is positive
or negative, the expected ground state is  singlet
superconductivity or a CDW. A simple calculation shows that for
$\vartheta<\vartheta_C$ the ground state is CDW and for larger
filling ($\vartheta>\vartheta_C$) we expect superconductivity,
where $\cos2\vartheta_C=-1/5$ (this corresponds to filling
$n_C=0.564$). On the other hand, if the exponent
$2\lambda_3-\lambda_1$ is larger, then $g_1-2g_2$ is always
positive. Since $g_1$ is positive, we get triplet
superconductivity. The two exponents are equal if
$U\sin^2\varphi=-2U'\cos^2\varphi\cos 2\vartheta$. The phase
diagram for this model is then shown in
Fig.~\ref{fig:phase_1d}(a), where we can see that the nature of
superconductivity changes as the Coulomb interaction increases
between the heavy orbital and light orbital on the neighboring
site.

There is also a strong mass renormalization in this model.
It is easy to see, that $\chi_1=\chi_2=\chi_3=\chi$ are equal
[see Eq.~(\ref{eq:def_chi})], and
\begin{equation}
   \chi\sim\tilde t_1^2+\tilde t_2^2+\tilde t_3^2+\tilde t_4^2 \;.
\end{equation}
 Putting in the actual expression for $\tilde t$-s, we get
\begin{equation}
 \chi = (2\tilde t\cos^3\varphi)^2 \left\{ \begin{array}{ll}
       4\cos^2 \vartheta (\cos^4\vartheta +2 \sin^4 \vartheta) \;,\;
        &\mbox{if} \;\; \lambda_1>\lambda_3 \\
        \sin^4 2\vartheta \cos^2 \vartheta) \;,\;
        &\mbox{if}\;\; \lambda_1<\lambda_3 \quad . \\
  \end{array} \right.
\end{equation}
 which become large for intermediate fillings.

 If the parity of the light and heavy orbitals is not the same
(e.g. light $s-$ or $d-$ orbitals and heavy $f$ orbitals, see
Fig.~\ref{fig:1Dmod}(b)), then
the assisted  hopping is odd and
\begin{equation}
  \tilde t_\delta =
    \left\{
      \begin{array}{ll}
        0  \;,\; &\mbox{if}\;\; \delta=0 \\
        \delta\tilde t \;,\; &\mbox{if}\;\; \delta=\pm 1 \quad . \\
      \end{array}
    \right.
\end{equation}
The parity of the assisted hopping has no effect to the
eigenvalues and eigenvectors of the ${\sf UF}$. Using
Eq.~(\ref{eq:calc_tijk0}), the
$\tilde t^{(0)}_{ijk}$-s are in this case
\begin{eqnarray}
  \tilde t^{(0)}_{333} & = & { \tilde t \over 4} \;,
    \nonumber\\
\tilde t^{(0)}_{113} & = & \tilde t^{(0)}_{131} = \tilde t^{(0)}_{113} =
  {\tilde t } {1\over \lambda^2_1 }\cos^2 \vartheta \cos^2 \varphi  \;,
\end{eqnarray}
and for the effective assisted hopping we get
\begin{eqnarray}
 \tilde t(\kappa_1,\kappa_2,\kappa_3;\omega)  = -2i
  \tilde t \cos^3\varphi
    &&\left\{
       \bigl[ \sin \kappa_1 \vartheta\sin \kappa_2 \vartheta
              \sin \kappa_3  \vartheta + \cos^2 \vartheta
          (\sin \kappa_1 \vartheta+\sin \kappa_2 \vartheta )
          \bigr]  \bigl( D / \omega \bigr)^{\rho \lambda_3}
    \right. \nonumber \\
    && -  \left.
        \cos^2\vartheta \sin \kappa_3 \vartheta
         \bigl( D / \omega \bigr)^{\rho (2\lambda_1-\lambda_3)}
       \right\} \;.
\end{eqnarray}
In this case the $\tilde t_1$, $\tilde t_2$, $\tilde t_3$ and $\tilde t_4$ are
\begin{eqnarray}
  \tilde t^*_1 &=& \tilde t_2 = -2i \tilde t \cos^3\varphi
    \left[
     \sin^3 \vartheta \bigl( D / \omega \bigr)^{\rho \lambda_3}
      + \cos^2 \vartheta \sin \vartheta
     \bigl( D / \omega \bigr)^{\rho(2\lambda_1-\lambda_3)}
    \right] \nonumber \\
 \tilde t_3 &=& -2i\tilde t \cos^3\varphi
    \left[
       \bigl(
          -\sin^3 \vartheta + 2 \cos^2 \vartheta \sin \vartheta
       \bigr)
       \bigl( D / \omega \bigr)^{\rho\lambda_3}
       + \cos^2 \vartheta \sin \vartheta
         \bigl( D / \bigr)^{\rho(2\lambda_1-\lambda_3)}
   \right] \nonumber \\
 \tilde t_4 &=& -2i\tilde t \cos^3\varphi
   \left[
       \bigl( \sin^3 \vartheta + 2 \cos^2 \vartheta \sin \vartheta
       \bigr)  \bigl( D / \omega \bigr)^{\rho\lambda_3}
       - \cos^2 \vartheta \sin \vartheta
         \bigl( D /\omega \bigr)^{\rho(2\lambda_1-\lambda_3)}
    \right] \nonumber \\
\end{eqnarray}
Comparing them to those of the even assisted hopping [Eqs.~\ref{eq:1Devent}],
we can see, that disregarding the prefactor of $i$ and the
exponents, the only difference is the interchange of $\sin\vartheta$ and
$\cos\vartheta$.

 Keeping only the most divergent parts in the assisted hopping, for the
induced interaction we get
\begin{equation}
 g_2 = (2\tilde t\cos^3\varphi)^2
   \left\{
     \begin{array}{ll}
       0 \;,\;
        &\mbox{if}\;\; \lambda_1>\lambda_3 \\
       -2\sin^2 2\vartheta \cos 2 \vartheta
         (D/\varepsilon_h)^{2\lambda_1}\;,\;
        &\mbox{if}\;\; \lambda_1<\lambda_3 \quad . \\
     \end{array}
   \right.
\end{equation}
If $\lambda_1<\lambda_3$, then $g_2$ is negative for small fillings and became
positive for larger fillings.

For the backward scattering
\begin{equation}
 g_1 = (2\tilde t\cos^3\varphi)^2
  \left\{
    \begin{array}{ll}
      \sin^2 2\vartheta (1+\cos 2\vartheta)
      (D/\varepsilon_h)^{2\lambda_3}\;,\;
      &\mbox{if}\;\; \lambda_1>\lambda_3 \\
      -\sin^2 2\vartheta (1-\cos 2\vartheta)
      (D/\varepsilon_h)^{4\lambda_1-2\lambda_3} \;,\;
      &\mbox{if}\;\; \lambda_1<\lambda_3 \quad . \\
    \end{array}
  \right.
\end{equation}
The boundary between the two region with different exponent is
the same as it was in the case of even $\tilde t_{\bbox{\delta}}$. If
$\lambda_1>\lambda_3$, both $g_1$ and $g_1-2g_2$ are positive,
so the ground state is TS. In the region
where $\lambda_1<\lambda_3$ holds, the $g_1$ is negative, and
for $\vartheta<\vartheta_C$ we get triplet superconductivity
and for $\vartheta>\vartheta_C$ the ground state is CDW. Here
$\vartheta$ is defined as $\cos 2\vartheta_C=1/5$ and the
corresponding filling is $\nu_C=0.436$ . The phase diagram is shown
in Fig.~\ref{fig:phase_1d}b.

We can conclude that the main effect of changing the parity leads to
exchange of the CDW and superconducting ground state.

In the present calculation we have neglected the interaction
between the light electrons.  That interaction changes the $g$-s, so
that the phase diagram changes also and new phases appear
near $\nu=0$ and $\nu=1$.

\section{Two-dimensional square lattice: $C\lowercase{u}O_2$ plane}

In the compounds characterized by $CuO_2$ planes a possible
representation of the $h$- orbitals are the two non-bonding
$p$-orbitals on the apical oxygens located below or above the
$Cu$ sites in the $CuO_2$ plane (the $O^{2-}$ ions are in the
tetrahedral position around $Cu^{2+}$ ion). For details see
Ref.~\onlinecite{Za89,Za89Nob}.
The $CuO_2$ plane forms a two dimensional square lattice
depicted on Fig.~\ref{fig:2Dmod}, where the $h$-orbitals are at
the corners and the $l$-band is formed by the orbitals $a_x$,
$a_y$ and $b$ at the middle of the sides and at corners,
respectively. One possible choice of the symmetry of the
orbitals corresponds to the $CuO_2$ plane with $p_x\>(a_x)$ ,
$p_y\>(a_y)$ and $d_{x^2-y^2}\>(b)$, while the two heavy
labelled by $\gamma=x,y$ are of $p_x$- and $p_y$-type .  For
this model the hopping to heavy orbitals is
\begin{equation}
  t^\gamma_{h,{\bbox{\delta}}} = t_h P^\gamma_{{\bbox{\delta}}} \nonumber\\
\end{equation}
and the hoppings between light orbitals are
\begin{equation}
  t_{{\bbox{\delta}}{\bbox{\delta}}'} =
    \left\{
      \begin{array}{ll}
       t p_{{\bbox{\delta}}}  \;,\; &\mbox{if}\;\; {\bbox{\delta}}'=(0,0) \\
       t' p_{{\bbox{\delta}}{\bbox{\delta}}'} \;,\; &\mbox{otherwise} \;, \\
      \end{array}
    \right.
\end{equation}
furthermore,
\begin{equation}
  \varepsilon_{{\bbox{\delta}}} =
    \left\{
      \begin{array}{ll}
        \varepsilon_b  \;,\; &\mbox{if}\;\; {\bbox{\delta}}'=(0,0) \\
        0 \;,\; &\mbox{otherwise} \; . \\
       \end{array}
    \right.
\end{equation}
The $p^{\vphantom{\dagger}}_{{\bbox{\delta}}}$ ,
$p^{\vphantom{\dagger}}_{{\bbox{\delta}},{\bbox{\delta}}'}$ and
$P^\gamma_{{\bbox{\delta}}}$
stand for the relative signs of the real wave functions in the overlap
regions. For the nearest neighbor hopping between oxygen and copper in the
$CuO_2$ plane
$p^{\vphantom{\dagger}}_{{\bbox{\delta}}}
=-p^{\vphantom{\dagger}}_{-{\bbox{\delta}}}$ holds and
for the next to the nearest neighbor hopping between oxygens
$p_{{\bbox{\delta}},{\bbox{\delta}}'}=\pm 1$ with $+(-)$ sign if
${\bbox{\delta}}-{\bbox{\delta}}'$  is parallel (perpendicular) to the (1,1)
direction
and for the apical oxygen
$P^\gamma_{\bbox{\delta}}=P^\gamma_{-{\bbox{\delta}}}$ if
${\bbox{\delta}}$ is parallel with $\gamma=x,y$ and zero otherwise.

 The Coulomb repulsion is
\begin{equation}
  U^\gamma_{{\bbox{\delta}}} =  \left\{ \begin{array}{ll}
       U  \;,\; &\mbox{if}\;\; {\bbox{\delta}}=0 \\
       U' \;,\; &\mbox{otherwise} \quad . \\
  \end{array} \right.
\end{equation}
where $U$ is the repulsion between electrons on the apex-$O$
and $Cu$, while $U'$ is between apex-$O$ and the next oxygens in the plane.

The assisted hopping is given by
\begin{equation}
  \tilde t^\gamma_{{\bbox{\delta}}} = \tilde t P^\gamma_{{\bbox{\delta}}} \;.
\end{equation}

In the following we will consider two limiting cases.

\subsection{ Case of $U'=t'=0$}

Here we are considering the extreme case where both $t'$ and
$U'$ are zero.
The dispersion relation is given by
\begin{equation}
  \omega_\pm = {1\over 2}
  \left(
     \varepsilon_b\pm\sqrt{\varepsilon_b^2+4{\mit\Omega}^2}
  \right) \;,
\end{equation}
where the $+(-)$ stands for upper(lower) band, ${\mit\Omega^2}=t^2_{xk}+
t^2_{yk}$ and $t_{\alpha k}=2t\sin k_\alpha a/2$  ($\alpha=x,y$).
The $\phi$-s [see Eq.~(\ref{eq:a=phi_d})] are given by
\begin{equation}
  \phi_0 = {\omega_\pm\over\sqrt{\omega_\pm^2+{\mit\Omega}^2}},
  \quad {\rm and} \quad
  \phi_\delta={t_{\delta k}\over\sqrt{\omega_\pm^2+{\mit\Omega}^2}} \;.
\end{equation}

The calculations of $\rho$ and $F$-s, defined by Eqs.~(\ref{eq:rho}) and
(\ref{eq:Fdd})], are straightforward:
\begin{equation}
  \rho = {1\over(2\pi)^2}
    { \sqrt{\varepsilon_b^2+4{\mit\Omega}^2} \over t^2} {\rm K}'(c) \;,
\end{equation}
and
\FL
\begin{eqnarray}
  F_{00} & = & |\phi_0|^2 \nonumber\\
  F_{x0} & = & F_{y0} = F^*_{-x0} = F^*_{-y0} =
     {i\over 2} \left(2t\over\omega_\pm\right) (1-c)
    |\phi_0|^2 \nonumber\\
  F_{xx} & = & F_{-x,-x} = F_{yy}=F_{-y,-y}=
     {1\over 2}\left(2t\over\omega_\pm\right)^2
    (1-c) |\phi_0|^2 \nonumber\\
  F_{xy} & = & -F_{x,-y}=-F_{-xy}=F_{-x,-y}=
    {1\over 2}\left(2t\over\omega_\pm\right)^2
    \left[
       {{\rm E}'(c)\over {\rm K}'(c)} -c
    \right] |\phi_0|^2 \nonumber\\
  F_{x,-x} & = & F_{-x,x}=F_{y,-y}=F_{-y,y}=
      {1\over 2}\left(2t\over\omega_\pm\right)^2
    \left[
      c-1 -2 c^2 + 2{{\rm E}'(c)\over {\rm K}'(c)}
    \right] |\phi_0|^2 \;,
\end{eqnarray}
where $c=1-{\mit \Omega}^2/(2t)^2$ and the elliptic function are defined as
\begin{equation}
  {\rm K}(k)=\int_0^{\pi/2}{d\varphi\over \sqrt{1-k^2\sin^2\varphi}}
  \quad {\rm and} \quad
  {\rm E}(k)=\int_0^{\pi/2}d\varphi \sqrt{1-k^2\sin^2\varphi}
\end{equation}
and ${\rm K}'(k)={\rm K}(k')$, ${\rm E}'(k)={\rm E}(k')$ where
$k'=\sqrt{1-k^2}$.

The vectors ${\bf s}$ and ${\bf r}$ are [see Eq.~(\ref{eq:spectFU})]
\begin{equation}
 {\bf s}^{(1)} =
   \left[ \begin{array}{c}
          0   \\ 0  \\  0  \\ 0  \\ 1
       \end{array}\right] \:,\quad
 {\bf s}^{(2)} = \left[ \begin{array}{c}
          i  \\ -i  \\  0  \\ 0  \\-2f
    \end{array}\right] \:,\quad
 {\bf s}^{(3)} = \left[ \begin{array}{c}
          0  \\ 0  \\ i  \\ -i  \\-2f
     \end{array}\right] \quad
 {\bf s}^{(4)} = \left[ \begin{array}{c}
          1  \\ 1  \\  0  \\ 0  \\ 0
     \end{array}\right] \:,\quad
 {\bf s}^{(5)} = \left[ \begin{array}{c}
          0  \\ 0  \\  1  \\ 1  \\ 0
     \end{array}\right] \;,
\end{equation}
and
\begin{eqnarray}
 {\bf r}^{(1)} & = & ( -if  \:,\; if \:,\; -if \:,\; if \:,\; 1 )\;,
\nonumber\\
 {\bf r}^{(2)} & = & ( -i/2 \:,\; i/2 \:,\; 0  \:,\; 0  \:,\; 0 )\;,
\nonumber\\
 {\bf r}^{(3)} & = & (  0  \:,\; 0  \:,\; -i/2 \:,\; i/2 \:,\; 0 )\;,
\nonumber\\
 {\bf r}^{(4)} & = & (  1/2  \:,\; 1/2  \:,\;  0 \:,\; 0  \:,\; 0 )\;,
\nonumber\\
 {\bf r}^{(5)} & = & (  0 \:,\; 0 \:,\; 1/2  \:,\; 1/2  \:,\; 0 )
\;,
\end{eqnarray}
with eigenvalues $\lambda_1=U|\phi_0|^2$ and
$\lambda_2=\lambda_3=\lambda_4=\lambda_5=0$, and
\begin{equation}
  f = (1-c) t/ \omega_\pm \;,
\end{equation}
furthermore the $\bbox{\delta}$ indices are ordered as $(x,-x,y,-y,0)$.
The $\xi$ functions [see Eq.~(\ref{eq:def_xi})] are
\begin{eqnarray}
 \xi_1(\bbox{\kappa}) & = & \phi_0 \;,\nonumber\\
 \xi_2(\bbox{\kappa}) & = & 2{2t \over \omega_\pm} \phi_0
      \sin^2(k^{\vphantom{\dagger}}_{Fx}(\bbox{\kappa}) a /2)- 2 f
\phi_0 \;,\nonumber\\
 \xi_3(\bbox{\kappa}) & = & 2{2t \over \omega_\pm} \phi_0
      \sin^2(k^{\vphantom{\dagger}}_{Fy}(\bbox{\kappa}) a /2) - 2 f
\phi_0 \;,\nonumber\\
 \xi_4(\bbox{\kappa}) & = & {2t \over \omega_\pm} \phi_0
      \sin(k^{\vphantom{\dagger}}_{Fx}(\bbox{\kappa}) a ) \;,\nonumber\\
 \xi_5(\bbox{\kappa}) & = & {2t \over \omega_\pm} \phi_0
      \sin(k^{\vphantom{\dagger}}_{Fy}(\bbox{\kappa}) a )\;.
\end{eqnarray}

The nonvanishing averages of $\xi\xi^*$
functions are from Eq.~(\ref{eq:xixi})
\begin{eqnarray}
 \langle|\xi_1(\bbox{\kappa})|^2\rangle  & = & |\phi_0|^2 \;,\nonumber\\
 \langle|\xi_4(\bbox{\kappa})|^2\rangle  & = &
    \langle|\xi_5(\bbox{\kappa})|^2\rangle  =
      2\left(2t\over\omega_\pm\right)^2
    \left(
      {{\rm E}'(c)\over {\rm K}'(c)} - c^2
    \right) |\phi_0|^2 \;.
\end{eqnarray}

Calculating the renormalized assisted hopping, we pick up the
most divergent term in Eq.~(\ref{eq:sol_tkkk}). The largest
value of the exponent $\lambda_i+\lambda_j-\lambda_k$
corresponds to the choice of $i=j=1$ and, since the remaining
four eigenvalues are degenerate, with $k=$2,3,4 or 5.
We can determine the initial values $\tilde t^{\gamma(0)}$
from Eq.~(\ref{eq:calc_tijk0}), and we get
$t^{x(0)}_{114} = -\tilde t f^2 $ and
$t^{y(0)}_{115} = -\tilde t f^2 $ with other
$t^{\gamma(0)}_{11k}$-s being equal to 0. So the assisted
hopping is
\begin{equation}
 t^x(\bbox{\kappa}_1,\bbox{\kappa}_2,\bbox{\kappa}_3;\omega) =
    t^{x(0)}_{114} \xi_1(\bbox{\kappa}_1)
    \xi_1(\bbox{\kappa}_2)\xi^*_4(\bbox{\kappa}_3)
  \left( D \over \omega \right)^{2\lambda_1-\lambda_4}
\end{equation}
and for $ t^y(\bbox{\kappa}_1,\bbox{\kappa}_2,\bbox{\kappa}_3;\omega)$
the
$\xi^*_4(\bbox{\kappa}_3)$ is replaced by $\xi^*_5(\bbox{\kappa}_3)$, so
that
\begin{eqnarray}
 t^\gamma(\bbox{\kappa}_1,\bbox{\kappa}_2,\bbox{\kappa}_3;\omega) =
  -\tilde t f^2 |\phi_0|^2 \phi_0 {2t \over \omega_\pm}
   \sin(k^{\vphantom{\dagger}}_{F\gamma}(\bbox{\kappa}_3) a)
   \left( D \over \omega \right)^{2 U |\phi_0|^2 \rho} \;,
\end{eqnarray}
where ${\bf k}_F=(k_{Fx},k_{Fy})$ is the Fermi vector parallel
to the $\bbox{\kappa}$.

 The expressions for $\chi$-s, according to Eq.~(\ref{eq:def_chi}), are
\begin{eqnarray}
  \chi_1 = \chi_2 &=&
   t^{x*}_{114}t^{x}_{114}
     |\xi_1|^2 \bigl\langle |\xi_1|^2 \bigr\rangle
     \bigl\langle |\xi^*_4(\bbox{\kappa})|^2 \bigr\rangle
   + t^{y*}_{115} t^{y}_{115}
     |\xi_1|^2 \bigl\langle |\xi_1|^2 \bigr\rangle
     \bigl\langle |\xi^*_5(\bbox{\kappa})|^2 \bigr\rangle
  \nonumber\\
  \chi_3(\bbox{\kappa}) & = &
     t^{x*}_{114}t^{x}_{114} \bigl\langle |\xi_1|^2 \bigr\rangle
     \bigl\langle |\xi_1|^2 \bigr\rangle |\xi^*_4(\bbox{\kappa})|^2
   + t^{y*}_{115} t^{y}_{115} \bigl\langle |\xi_1|^2 \bigr\rangle
   \bigl\langle |\xi_1|^2 \bigr\rangle |\xi^*_5(\bbox{\kappa})|^2 \;,
\end{eqnarray}
or, inserting  the $\xi$-s and $t^{\gamma(0)}_{ijk}$-s,
\begin{eqnarray}
  \chi_1 & = & \chi_2 = {1 \over 16}\phi_0^6
  {{\mit \Omega}^8 \over \omega_\pm^6t^2}
  \left[
    { {\rm E}'(c)\over {\rm K}'(c)}-c^2 \right] \tilde t^2
  \nonumber\\
  \chi_3(\bbox{\kappa}) & = & {1 \over 64}
  \phi_0^6 {{\mit \Omega}^8 \over \omega_\pm^6t^2}
  \bigl[ \sin^2(k^{\vphantom{\dagger}}_{Fx}(\bbox{\kappa})a)
    + \sin^2(k^{\vphantom{\dagger}}_{Fy}(\bbox{\kappa})a)
  \bigr]
  \tilde t^2 \;.
\end{eqnarray}

A straightforward calculation gives the interaction in the
$S$-wave channel,
\begin{equation}
  V^S(\bbox{\kappa},-\bbox{\kappa},-\bbox{\kappa}',\bbox{\kappa}')
    =-2\rho\chi A(D/\varepsilon_h) \;.
\end{equation}
Since the interaction is independent of momentum, the
$\bbox{\kappa}$ dependence of the
superconducting order parameter $\Delta^S$ is small [due to
small $\bbox{\kappa}$ dependence of $Z(\bbox{\kappa})$, see
Eq.~(\ref{eq:Tc_selfcons_int})].
The approximate strength of the dimensionless
effective coupling in the singlet channel [see Eq.~(\ref{eq:gseff})] is
\begin{equation}
  g_{eff}^{(s)}\approx -
  {2 \rho^2 \chi A(D/\varepsilon_h)
   \over
   1+3\chi \rho A(D/\varepsilon_h)} \;,
\end{equation}
where we can see immediately that
\begin{equation}
  -g_{eff}^{(s)}<{2\over 3} \;.
\end{equation}
The relation of the transition temperature $T_c$ and the mass enhancement is
given by the curve $q=2$ in Fig.~\ref{fig:tc}. The highest $T_c$ is obtained
for mass enhancement about 2.5--3.5.

  Similar calculation for the triplet channel gives
\begin{eqnarray}
  V^T&&(\bbox{\kappa},-\bbox{\kappa},-\bbox{\kappa}',\bbox{\kappa}')=
  {1\over 32} \tilde t^2 |\phi_0|^6 \rho {\Omega\over t^2 \omega_\pm^6}
  A(D/\varepsilon_h) \nonumber\\
  &&\times
   \left[
    \sin(k^{\vphantom{\dagger}}_{Fx}(\bbox{\kappa})a)
    \sin(k^{\vphantom{\dagger}}_{Fx}(\bbox{\kappa}')a) +
    \sin(k^{\vphantom{\dagger}}_{Fy}(\bbox{\kappa})a)
    \sin(k^{\vphantom{\dagger}}_{Fy}(\bbox{\kappa}')a)
  \right] \;,
\end{eqnarray}
which results in repulsion.

\subsection{The case of $U=t=0$}

The tight binding Hamiltonian (\ref{eq:H0}) with the choice of parameters
corresponding to this case has the following form in the
${\bf k}$ representation:
\begin{equation}
  H=4t'\sum_{{\bf k},\sigma}\sin(k_xa/2) \sin(k_ya/2)
  \left(a^\dagger_{x,{\bf k},\sigma}a^{\vphantom{\dagger}}_{y,{\bf
k},\sigma}+{\rm h.c.}
\right) \;
  \label{eq:H0YBCO'}
\end{equation}
and can be diagonalized by introducing the
\begin{eqnarray}
  a_{x,{\bf k}}& = & \phi_x({\bf k})\left(d_{{\bf k},+}+d_{{\bf k},-}\right)
    \nonumber\\
  a_{y,{\bf k}}& = & \phi_y({\bf k})\left(d_{{\bf k},+}-d_{{\bf k},-}\right)
\end{eqnarray}
operators, where
$\phi_\alpha({\bf k})=\mbox{sign}(\sin k_\alpha a/2)/\sqrt{2}$
and the diagonalized Hamiltonian is
\begin{equation}
  H=4t'\sum_{{\bf k},\sigma}
  | \sin( k_xa/2)  \sin(k_ya/2) |
  \left(
      d^\dagger_{{\bf k},\sigma,+}d^{\vphantom{\dagger}}_{{\bf k},\sigma,+}
     -d^\dagger_{{\bf k},\sigma,-}d^{\vphantom{\dagger}}_{{\bf k},\sigma,-}
  \right)
  \;.
\end{equation}
 With this choice of $\phi_\delta$-s, the operators
 $d^\dagger_{{\bf k},\sigma,+}$ and
 $d^{\vphantom{\dagger}}_{{\bf k},\sigma,+}$
are
associated with the upper band which is combined from the upper parts
of the two bands labelled by $x$ and $y$. In the following, only the upper band
is kept and the
index + will be dropped.

For convenience, the momentum is shifted as $k_x\rightarrow k_x-\pi/a$
and $k_y\rightarrow k_y-\pi/a$ in all of the following formulas.
For example, the dispersion relation changes to
$\varepsilon=4t'\cos k_xa/2 \cos k_ya/2$. That shift, however,
can be completely incorporated by the amplitudes
$\phi_{\bbox{\delta}}(\bbox{\kappa})$:
\begin{equation}
  \phi_{\pm x}=\phi_{\pm y}=\pm i/\sqrt{2}
\end{equation}
in Eqs.~(\ref{eq:asst_unren}) and (\ref{eq:def_tddd}).

Introducing the notation $ F_{{\bbox{\delta}}{\bbox{\delta}}}=\tilde F_0$,
$ F_{{\bbox{\delta}},-{\bbox{\delta}}}=\tilde F_2$ and
$ F_{xy}=F_{-x-y}=-F_{-xy}=-F_{x-y}=\tilde F_1$, the matrix
$\tilde F_{{\bbox{\delta}}{\bbox{\delta}}'}$ takes the form
\begin{equation}
  {\sf F} =
  \left(
    \begin{array}{cccc}
      \tilde F_0&\tilde F_2&\tilde F_1&-\tilde F_1 \\
      \tilde F_2&\tilde F_0&-\tilde F_1&\tilde F_1 \\
      \tilde F_1&-\tilde F_1&\tilde F_0&\tilde F_2 \\
      -\tilde F_1&\tilde F_1&\tilde F_2&\tilde F_0 \\
    \end{array}
  \right),
\end{equation}
which eigenvectors and eigenvalues are easy to obtain:
\begin{equation}
 {\bf s}^{(1)} =
   \left[ \begin{array}{c}
          1   \\ -1  \\  1  \\ -1
       \end{array}\right] \quad
 {\bf s}^{(2)} = \left[ \begin{array}{c}
          1  \\ 1  \\  0  \\ 0
    \end{array}\right] \quad
 {\bf s}^{(3)} = \left[ \begin{array}{c}
          0  \\ 0  \\ 1  \\ 1
     \end{array}\right] \quad
 {\bf s}^{(4)} = \left[ \begin{array}{c}
          1  \\ -1  \\ -1  \\ 1
     \end{array}\right] \quad
\end{equation}
and
\begin{eqnarray}
 {\bf r}^{(1)} = {\bf s}^{(1)T}/4 \:,\:
 {\bf r}^{(2)} = {\bf s}^{(2)T}/2 \:,\:
 {\bf r}^{(3)} = {\bf s}^{(3)T}/2 \:,\:
 {\bf r}^{(4)} = {\bf s}^{(4)T}/4 \:,\:
\end{eqnarray}
where the indices ${\bbox{\delta}}$ are ordered as $(x,-x,y,-y)$ and the
superscript denotes a transposed vector.

The corresponding eigenvalues are
\begin{eqnarray}
  \lambda_1 &=& U' (\tilde F_0 -\tilde F_2 +2\tilde F_1)\;, \nonumber\\
  \lambda_2 &=& \lambda_3 = U' (\tilde F_0 +\tilde F_2)\;, \nonumber\\
  \lambda_4 &=& U' (\tilde F_0 -\tilde F_2 -2\tilde F_1)\;.
\end{eqnarray}

The integrals in $\tilde F_0$, $\tilde F_1$ and $\tilde F_2$ leads to
elliptic functions
\begin{eqnarray}
  \rho & = &
   {\rm K'}(\varepsilon^{\vphantom{\dagger}}_F/4t')/\pi^2 t'
  \nonumber\\
  \tilde F_0  & = & \left\langle |\phi_x|^2 \right\rangle
    = 1/2
  \nonumber\\
  \tilde F_1 & = &
    \left\langle \phi^*_x\phi_y \cos(k_xa/2)\cos(k_ya/2)\right\rangle
   = \varepsilon^{\vphantom{\dagger}}_F / 8t'
   \nonumber\\
  \tilde F_2 & = &
    \left\langle |\phi_x|^2 \cos k_xa  \right\rangle
    =1/2 -
       {\rm E}'(\varepsilon^{\vphantom{\dagger}}_F/4t')
      /{\rm K}'(\varepsilon^{\vphantom{\dagger}}_F/4t')
    \;.
\end{eqnarray}
Furthermore, using Eq.~(\ref{eq:def_xi}) we get
the functions $\xi$:
\begin{eqnarray}
 \xi_1(\bbox{\kappa}) & = & -i \sqrt{2} \left[
        \cos(k^{\vphantom{\dagger}}_{Fx}(\bbox{\kappa}) a /2)
       +\cos(k^{\vphantom{\dagger}}_{Fy}(\bbox{\kappa}) a /2)
   \right]
\nonumber\\
 \xi_2(\bbox{\kappa}) & = &
   - \sqrt{2} \sin(k^{\vphantom{\dagger}}_{Fx}(\bbox{\kappa}) a /2)
\nonumber\\
 \xi_3(\bbox{\kappa}) & = & -\sqrt{2}
\sin(k^{\vphantom{\dagger}}_{Fy}(\bbox{\kappa}) a /2)
\nonumber\\
 \xi_4(\bbox{\kappa}) & = & -i \sqrt{2} \left[
     \cos(k^{\vphantom{\dagger}}_{Fx}(\bbox{\kappa}) a /2)
      -\cos(k^{\vphantom{\dagger}}_{Fy}(\bbox{\kappa}) a /2)
  \right] \;,
\end{eqnarray}
and the nonvanishing averages of $\xi\xi^*$ are:
\begin{eqnarray}
 \langle|\xi_1(\bbox{\kappa})|^2\rangle  & = &
   4 ( \tilde F_0 - \tilde F_2 + 2\tilde F_1)=4\lambda_1
 \nonumber\\
 \langle|\xi_2(\bbox{\kappa})|^2\rangle  & = &
 \langle|\xi_3(\bbox{\kappa})|^2\rangle   =
   2 ( \tilde F_0 + \tilde F_2 )=2\lambda_2
 \nonumber\\
 \langle|\xi_4(\bbox{\kappa})|^2\rangle  & = &
   4 ( \tilde F_0 - \tilde F_2 - 2\tilde F_1)=4\lambda_4 \;.
\end{eqnarray}

The leading terms in Eq.~(\ref{eq:sol_tkkk}) with proper
symmetry are those with the exponents $2\lambda_1-\lambda_2$ or
$\lambda_1+\lambda_2-\lambda_4$,
and the corresponding $\tilde t^{\gamma(0)}_{ijk}$ are
\begin{equation}
  \tilde t_{112}^{x(0)} = \tilde t_{113}^{y(0)} = \tilde t'/ 16 \;,
\end{equation}
or
\begin{equation}
  \tilde t_{124}^{x(0)}  = \tilde t_{214}^{x(0)} =-\tilde t_{134}^{y(0)} =
  -\tilde t_{314}^{y(0)} = \tilde t' / 16 \;,
\end{equation}
depending on whether
$\varepsilon^{\vphantom{\dagger}}_F/4t>0.41$ or $<0.41$ , respectively, so that
\begin{equation}
  \tilde t^x(\bbox{\kappa}_1,\bbox{\kappa}_2,\bbox{\kappa}_3)=
  {\tilde t' \over 16}
  \xi_1(\bbox{\kappa}_1)\xi_1(\bbox{\kappa}_2)
  \xi_2^*(\bbox{\kappa}_3)
  \left( D \over \omega \right)^{(2\lambda_1-\lambda_2)\rho}
  \label{eq:casebtkkk1}
\end{equation}
for $\varepsilon^{\vphantom{\dagger}}_F/4t'<0.41$
and
\begin{equation}
  t^x(\bbox{\kappa}_1,\bbox{\kappa}_2,\bbox{\kappa}_3)  =
    {\tilde t' \over 16}
      \left[
        \xi_1(\bbox{\kappa}_1) \xi_2(\bbox{\kappa}_2)
       +\xi_2(\bbox{\kappa}_1)\xi_1(\bbox{\kappa}_2)
      \right]
      \xi^*_4(\bbox{\kappa}_3)
    \left( D \over \omega \right)^{(\lambda_1+\lambda_2-\lambda_4)\rho}
    \label{eq:casebtkkk2}
\end{equation}
for $\varepsilon^{\vphantom{\dagger}}_F/4t'>0.41$ and for
$t^y(\bbox{\kappa}_1,\bbox{\kappa}_2,\bbox{\kappa}_3)$ the
$\xi_2(\bbox{\kappa}_2)$ should be replaced by $\xi_3(\bbox{\kappa}_2)$ and
multiplied by a minus sign for $\varepsilon^{\vphantom{\dagger}}_F/4t'>0.41$.

The $\chi$-s, defined by Eq.~(\ref{eq:def_chi}), calculated by using
Eqs.~(\ref{eq:sol_tkkk}) and (\ref{eq:casebtkkk1}-\ref{eq:casebtkkk2}) are
$\bbox{\kappa}$ dependent with
\begin{eqnarray}
  \tilde\chi_1(\bbox{\kappa}) & = &
    \tilde\chi_2(\bbox{\kappa}) = {\tilde  t^2 \over 16}
    \lambda_1\lambda_2 |\xi_1(\bbox{\kappa})|^2
 \nonumber \\
    \tilde\chi_3(\bbox{\kappa}) & = & {\tilde  t^2 \over 16}
    \lambda_1^2
\bigl(|\xi_2(\bbox{\kappa})|^2+|\xi_3(\bbox{\kappa})|^2\bigr)
  \nonumber
\end{eqnarray}
for $\varepsilon^{\vphantom{\dagger}}_F/4t'>0.41$ and
\begin{eqnarray}
  \tilde\chi_1(\bbox{\kappa}) & = &
    \tilde\chi_2(\bbox{\kappa}) = {\tilde  t^2 \over 16} \lambda_4
    \left[
      4\lambda_1
      \bigl(|\xi_2(\bbox{\kappa})|^2+|\xi_3(\bbox{\kappa})|^2 \bigr) +
      4 \lambda_2 |\xi_1(\bbox{\kappa})|^2 \right]
 \nonumber \\
    \tilde\chi_3(\bbox{\kappa}) & = & {\tilde  t^2 \over 8}
    |\xi_4(\bbox{\kappa})|^2
\end{eqnarray}
for $\varepsilon^{\vphantom{\dagger}}_F/4t'<0.41$.
The average of
$\tilde\chi_j(\bbox{\kappa})$-s over the Fermi surface is
\begin{equation}
  \tilde\chi=\left\{
  \begin{array}{ll}
    \lambda_2\lambda_1^2/4 \;,\;
         &\mbox{if}\;\; \varepsilon^{\vphantom{\dagger}}_F/4t'>0.41 \\
    \lambda_1\lambda_2\lambda_4/2 \;,\;
         &\mbox{if}\;\; \varepsilon^{\vphantom{\dagger}}_F/4t'<0.41
         \quad . \\
  \end{array}\right.
\end{equation}

 In the case $\varepsilon^{\vphantom{\dagger}}_F/4t'<0.41$, for the singlet
superconductivity we get
\begin{equation}
  V^S(\bbox{\kappa},-\bbox{\kappa},-\bbox{\kappa}',\bbox{\kappa}')=
   -{\tilde t^2 \over  32} \rho \lambda_2
  |\xi_1(\bbox{\kappa})|^2 |\xi_1(\bbox{\kappa}')|^2 A(D/|\varepsilon_h|)
\;,
\end{equation}
which gives rise to an $S$-type superconductivity.
The triplet spin channel contribution a repulsive interaction of a $p$-type.

 For the case $\varepsilon^{\vphantom{\dagger}}_F/4t'>0.41$ we get a $d$-wave
repulsive interaction in the singlet channel.(In Ref.~\onlinecite{ZaPeZi} we
got $d$--wave attraction, which was due to a sign error.)

\section{Discussion}
The role of the assisted hopping is demonstrated in models where
additional to the conduction band there are orbitals near the
Fermi surface. The
occupations of these orbitals fluctuate between two values. The
state of higher occupation is obtained by adding a heavy
electron . All the other states are ignored. The energies of
these two states include all of the intratomic Coulomb
interaction, thus these states are fully renormalized in
the atomic sense. We call the attention to two physical realizations:

(i){\it Heavy $f$--electrons}.
 In this case it must be assumed that
one of the renormalized $f$--levels is near to the Fermi energy on
the scale of the Fermi energy. If for one of the $f$--levels that
condition is satisfied, then the model can be applied by
considering the conduction electron assisted hybridization of
the $4f$--electrons with the conduction band. The assisted
process is induced by the change in the occupation of
conduction band in the atomic orbital (tight--binding)
representation. It is a striking feature, that the very
large mass enhancement in the conduction band practically
eliminates the superconducting state. The moderate mass
enhancement can be correlated with the superconductivity
(see chapter VI.). The possible role of the Coulomb interaction
in conduction band has been considered also in Ref.
\onlinecite{ZaPeZi} by the slave--boson technique and it makes
the superconducting state even more favorable.

(ii){\it $CuO_2$ plane}. The strikingly flat parts of the electronic
band structure are due to the non--bonding oxygen
orbitals\cite{Massetc}. E.g. in the $YBa_2Cu_3O_{7-\delta}$ compound
it is either due to the apex oxygens taking the tetrahedral
positions above or below the $Cu$ site or it is generated by
the non--bonding $\pi$--orbitals of the oxygens in the $CuO_2$
plane, which are oriented
perpendicular to the plane in the $z$-direction. These states can
hybridize with the orbitals on $Cu$ only if the $CuO_2$ plane is distorted
and the $Cu$ and $O$ atoms form separate planes. Such a flat
band has been recently observed by experiment\cite{Der} and
also reproduced by band structure calculation\cite{Ande}. The
previous case might be related to the influence of the
distance of the apex oxygen from the $CuO_2$ plane on the
superconducting transition temperature\cite{MiCa}. See for
more detailed discussion Refs. \onlinecite{Za89Nob,ZaPeZi,Za89}.
The calculated induced electron--electron interaction is
momentum dependent due to the form factors in the electron
assisted hybridization, but the models treated in
Ref.~\onlinecite{ZaPeZi} the transition leads to always $s$--type of
superconductors.

The Hamiltonian for a definite system must be constructed in the
atomic orbital picture, thus the tight binding approximation is
used. For the sake of simplicity only the conduction band
crossing the Fermi energy is kept for the light particle, even
if the other broad bands could contribute also to those
integrals which have logarithmic character in the most simpler
approximation.

The vertex corrections to the conduction electron assisted
hopping between the light and heavy bands are determined. The
dependence on the occupations of the heavy orbitals (i.e. interaction
like $c^\dagger h^\dagger h h$) does not
contribute in the logarithmic approximation, thus it is not
taken into account. The assisted hopping vertices
$\tilde t$ are strongly renormalized by the local Coulomb
interaction $U$ between the heavy and light--particles. The
general formulation is presented in Chapter III., where the
number of different couplings are finite as the momentum
dependence appears only in the form factors which belong
to some certain class. The solution of these vertex equations
for $\tilde t$ can be very different depending on the model. In
the case of onsite Coulomb interaction $U$, the renormalization
by $U$ can reduce the structure in the formfactor by integrating
out the dependence on the momenta of the scattered electron. That
happens in case A in Chapter VIII. [see Eqs.~(8.16)-(8.17)]. On
the other hand, if the interaction $U'$ is the next neighbor
Coulomb interaction, then the form factors of the vertex corrections
are altering in a certain class, but the level of the structure
is never reduced, see case B in Chapter VIII. In the
one--dimensional model new coupling is generated, and as it is
discussed in Chapter VII the structure of the vertex equations
is mapped to those in a one--dimensional interacting electron
gas with the two couplings $g_1$, $g_2$, $g_3$ and $g_4$ known as the
$g$--ology\cite{So79}. The results obtained are generalization
of those in Ref. \onlinecite{Za89PR} and may be relevant in the
quasi--one dimensional organic conductors. The functional form of the
vertex corrections are always power functions of $\omega/D$, but
the exponents and this way the strengths of the enhancement are
very sensitive on the actual structure of the formfactors.

The large mass enhancement described in Chapter IV. and
calculated in Chapter VII. and VIII. is a quite general
consequence of the theory. That can be very large, its value,
however, is limited by the low energy infrared cutoff due to the
dispersion and energy of the heavy band.

The superconductivity is determined by very similar expressions
as the mass enhancement [see e.g. Eqs.~(\ref{eq:Z}) and (\ref{eq:gSeff})].
Usually
in other theories the similar expressions are related by Ward
identities, but here those can not be exploited as the relevant
quantities appear in different channels (zero sound and Cooper).
The single particle weight $Z$ in the Green's function are
playing crucial role in the strength of the electron--electron
interaction and the mass enhancement . The related expressions
of Chapter VI. and Fig.~\ref{fig:tc} represents quite general
relations and they may be relevant in other models as well.

In the case of one--dimensional models a much richer class of
susceptibilities are discussed and the phase diagrams in
Fig.~\ref{fig:phase_1d} contains spin density wave (SDW) and
charge density wave (CDW) and triplet and singlet superconductivity

Finally, it is worth to point out, that in the actual
calculation the electrons on the different heavy sites are not
correlated which is the consequence of our systemic logarithmic
approximation schema, but that can be lost beyond the approximation
applied. As far as the number of the excited heavy particle levels
at a given time is small, i.e. the dynamics occurs on a dilute
set of orbitals, the approximation applied is justified.

For any certain problem with conduction electron assisted
hopping between a heavy and a light orbital the model can be
treated in the general schema presented. The large vertex
corrections make it promising, that the weaker assisted hoppings
can play a determining role in some systems, even if their bare
amplitude is weaker then the Coulomb interaction\cite{ZaPeZi}.

\acknowledgments

One of the authors A. Z. would like to thank for fruitful discussions
with D. Pines, D. Einzel and G. Zim\'anyi and is expressing his
gratitude for the hospitality and support by the Science and
Technology Center for Superconductivity and the Physics
Department of University of Illinois at Urbana. K. P. would like
to thank to useful discussions with J. S\'olyom.
This work has been supported by Hungarian AKA 1-300-2-93-0-834
and OTKA 2979/91 and 7283/93 and by
Swiss National Science Foundation Grants 20-33964.92 and 2000-037653.93/1.

\begin{figure}[t]
  \caption{The model for orbitals in one dimensions are shown.
The circles represents the
light $s$ orbitals, and inside them are the heavy $d$ (a) or $p$ (b)
orbitals (the $f$ orbitals are not presented as the $p$ orbitals  have
the same odd overlap as the deep $f$ orbitals). The energy levels,
the hoppings and interactions between them are shown in (c).
The clear and shadowed areas indicate the opposite signs of the wave functions.
\label{fig:1Dmod}
}
\end{figure}

\begin{figure}[t]
  \caption{
  The $CuO$ plane of the $YBCO$ compounds is shown as an example of the model
in two dimensions. The light $O$ and $Cu$ orbitals are found on the
sides and on the corners of the squares, respectively. The two orbitals of the
 apex oxygen below the $Cu$ sites can play the role of the heavy orbitals.
 \label{fig:2Dmod}
}
\end{figure}

\begin{figure}[t]
  \caption{
(a) The bare assisted hopping vertex is shown where the
double line and light lines stand for heavy and light particles,
respectively. The wavy line denotes the assisted hopping. (b)
The bare Coulomb interaction is indicated by dashed lines.
  \label{fig:bare_tU}
}
\end{figure}

\begin{figure}
\caption{
(a) The Coulomb corrections are shown in second order. The
diagrams contribute by logarithms, but they cancel.
(b) The corrections appearing in vertex equation are shown by time
ordered diagrams where the assisted hopping is renormalized by the
Coulomb interaction.
  \label{fig:corr}
}
\end{figure}

\begin{figure}
\caption{ The contributions to the  light particle
self-energy are shown by time ordered skeleton diagrams.
\label{fig:self_ene}
}
\end{figure}

\begin{figure}
  \caption{
 The interaction between the light particles induced by
the assisted hopping is shown separately for the different channels: (a) the
spin parallel and (b) antiparallel channels. The diagrams are time
ordered.
\label{fig:gen_vertex}
}
\end{figure}

\begin{figure}
  \caption{
 The renormalization of the light electron dispersion
curve is shown in the neighborhood of the Fermi energy
$\varepsilon^{\protect\phantom{+}}_F$. The renormalization is essential
in the range around the Fermi energy characterized by the low
energy cutoff $\varepsilon_h$. The large mass enhancement occurs
in an energy range $Z\varepsilon_h$.
  \label{fig:disp_curve}
}
\end{figure}

\begin{figure}
\caption{
The selfconsistent equation for the gap $\Delta$ is illustrated.
\label{fig:gap_eq}
}
\end{figure}

\begin{figure}
\caption{
The critical temperature as a function of mass enhancement for
different values of $q$ [see Eq.~(\protect\ref{eq:gSeff})]
\label{fig:tc}
}
\end{figure}

\begin{figure}
\caption{
  The band dispersion of the 1D model.
  \label{fig:1Ddisprel}
}
\end{figure}

\begin{figure}
\caption{
  The prefactor of
$(2\tilde t \cos^3 \varphi)^2 (D/\varepsilon_h)^{2\alpha}$
of the effective interaction
 $g_1$, $g_2$ and   the $\tilde\chi(\protect\bbox{\kappa})$
as a function of different filling for the even case: (a) $\lambda_1>\lambda_3$
and (b)
 $\lambda_1<\lambda_3$.
\label{fig:geff1D}
}
\end{figure}

\begin{figure}
\caption{
  The phase diagram of the 1D model in case of (a) even and
  (b) odd assisted hopping.
  \label{fig:phase_1d}
}
\end{figure}

\end{document}